\documentclass{aa}   

\usepackage{graphicx}
\usepackage{txfonts}
\usepackage{color}
\usepackage{textcomp}  
\usepackage{hyperref}

\begin{document}

\title{Small glitches and other rotational irregularities of the Vela pulsar}

\author{C.~M. Espinoza\inst{1} 
\and
D.~Antonopoulou\inst{2,3} 
\and
R.~Dodson\inst{4}
\and 
M.~Stepanova\inst{1}
\and
A.~Scherer\inst{5}
}

\institute{Departamento de F\'isica, Universidad de Santiago de Chile {\sc (USACH)}, Estaci\'on Central, Santiago 9170124, Chile \\
\email{cristobal.espinoza.r@usach.cl}
\and
Nicolaus Copernicus Astronomical Center, Polish Academy of Sciences, ul. Bartycka 18, PL-00-716 Warsaw, Poland
\and
Jodrell Bank Centre for Astrophysics, School of Physics and Astronomy, The University of Manchester, Manchester M13 9PL, UK
\and
International Centre for Radio Astronomy Research, University of
Western Australia, Crawley, WA 6009, Australia
\and
Instituto de Astrof{\'i}sica, Pontificia Universidad Cat{\'o}lica de Chile, 782-0436 Macul, Santiago, Chile
}

\date{}

 
\abstract
{Glitches are sudden increases in the rotation rate $\nu$ of neutron stars, which are thought to be driven by the neutron superfluid inside the star.
The Vela pulsar presents a comparatively high rate of glitches, with 21 events reported since observations began in 1968.
These are amongst the largest known glitches (17 of them have sizes $\Delta\nu/\nu\geq10^{-6}$) and exhibit very similar characteristics. 
This similarity, combined with the regularity with which large glitches occur, has turned Vela into an archetype of this type of glitching behaviour. 
The properties of its smallest glitches, on the other hand, are not clearly established.
} 
{ We explore the population of small-amplitude, rapid rotational changes in the Vela pulsar and determine the rate of occurrence and sizes of its smallest glitches. 
This will help advance our understanding of the actual distribution of glitch sizes and inter-glitch waiting times in this pulsar, which has implications for theoretical models of the glitch mechanism. 
} 
{ High-cadence observations of the Vela pulsar were taken between 1981 and 2005 at the Mount Pleasant Radio Observatory. 
An automated systematic search was carried out that investigated whether a significant change of spin frequency $\nu$ and/or the spin-down rate $\dot{\nu}$ takes place at any given time. 
} 
{ We find two glitches that have not been reported before, with respective sizes $\Delta\nu/\nu$ of $(5.55\pm0.03)\times10^{-9}$ and $(38\pm4)\times10^{-9}$. The latter is followed by an exponential-like recovery with a characteristic timescale of $31$\,d. 
 In addition to these two glitch events, our study reveals numerous events of all possible signatures (i.e. combinations of $\Delta\nu$ and $\Delta\dot{\nu}$ signs), all of them small with  $|\Delta\nu|/\nu<10^{-9}$, which contribute to the Vela timing noise. 
} 
{ The Vela pulsar presents an under-abundance of small glitches compared to many other glitching pulsars, which appears genuine and not a result of observational biases. In addition to typical glitches, the smooth spin-down of the pulsar is also affected by an almost continuous activity that can be partially characterised by small step-like changes in $\nu$, $\dot{\nu,}$ or both. 
Simulations indicate that a continuous wandering of the rotational phase, following a red spectrum,  could mimic such step-like changes in the timing residuals.
}


\maketitle
%

\section{Introduction}

Neutron stars, often observed as pulsars, typically display a very stable rotation. 
In isolation, their rotational evolution is rather smooth, decelerating at a slow rate due to energy losses. 
Nonetheless, various dynamical processes affect this evolution, especially so in young neutron stars. 
The observed effects appear either as wandering of the rotational phase around the predictions of a simple rotational model \citep[timing noise, see e.g. ][]{hlk10,psj+19} or as sudden accelerations of the rotation, called glitches \citep[$\Delta\nu>0$, where $\nu$ is the spin frequency of the star, see e.g.][]{elsk11}. 
Some glitches, often the largest ones, are accompanied by a step-like increase in the spin-down rate ($\Delta\dot{\nu}<0$), which then evolves towards pre-glitch values in a slow (tens to hundreds of days) process known as the glitch recovery \citep[][]{sl96,ymh+13}. 
It is unclear whether the two rotational phenomena have a similar origin or if different mechanisms are at play. 
The current consensus is that glitches are caused by an internal neutron superfluid component \citep{ai75,hm15}, but the origin of the timing noise is less clear and a variety of processes are still being considered \citep[e.g. magnetospheric fluctuations or superfluid turbulence;][]{klo+06,mw09,lhk+10,ml14}. 
We review 2.5 decades of timing observations of the Vela pulsar with the aim of characterising its rotational irregularities. 
This first article focuses on data features for which the distinction between timing noise and small glitches becomes less clear.

The Vela pulsar is a young and nearby neutron star associated with the Vela supernova remnant \citep[$\geq 10$\,kyr old;][]{aet95,tsk+09,sh14}.
It is the brightest radio pulsar in the southern sky, and its rotation ($\nu\approx11.2$\,Hz) has been extensively monitored since it was discovered in 1968 \citep{lvm68}.
Vela is the first pulsar that was found to glitch \citep{rm69,rd69}, and when a new glitch was detected two and a half years later, it became clear that its glitches are particularly large and frequent. The typical size of the spin frequency increase at the moment of a glitch, $\Delta\nu$, is about $20\,\mu$Hz. 
This is at the high end of the overall glitch size distribution, all pulsars considered,  which extends from $\sim10^{-4}$ to nearly $100\,\mu$Hz \citep{apj17,fer+17}. 
Moreover, these large glitches interrupt the rotation of Vela every $\sim3$\,yr, a rather high rate compared to most pulsars, which exhibit fewer than one such event every 10 years \citep[][]{elsk11}.

Observations of large glitches provide constraints to models of the internal structure of neutron stars and to the interaction between the crust and the internal neutron superfluid \citep[e.g.][]{lel99,hps12,aghe12,cha13,dcg+16}.
Furthermore, a collection of large glitches can be used to constrain superfluid properties and calculate the pulsar mass \citep{heaa15,pahs17}.
Whilst large glitches are easy to spot in timing data, small glitches ($\Delta\nu<1\,\mu$Hz) can be missed or be indistinguishable from timing noise when the cadence of the observations is not high enough or when the timing sensitivity is low  \citep{eas+14,slk+16}. 
By exploring the population of small-size irregularities, we can improve our understanding of both glitches and timing noise, and investigate whether they are connected. 
A study like this was carried out for the Crab pulsar and concluded that events with a typical glitch signature appear mostly beyond a certain $\Delta\nu$ size, whilst a separate, larger population of timing noise-like events emerges at smaller scales \citep{eas+14}.  Such a cutoff can have implications on glitch models and on the glitch trigger mechanism \citep{has16,aa18}.

Although the Vela pulsar has been extensively observed for more than four decades by several radio observatories \citep[e.g.][]{dlm07,buc13b}, there is no record of thorough searches aiming at generating complete catalogues of glitches in a given period of time. 
Perhaps the most relevant analysis in this respect is the detailed study by \citet{cdk88}, who tried to identify all glitches between November 1968 and March 1983.
They identified two types of events: micro- and macro-glitches, the latter corresponding to what we call glitches today: a positive step in frequency ($\Delta\nu>0$) together with a negative step in spin-down rate ($\Delta\dot{\nu}<0$), often followed by an exponential-like relaxation.
Micro-glitches, on the other hand, come in all sign combinations of $\Delta\nu$ and $\Delta\dot{\nu}$ and have much smaller amplitudes ($\Delta\nu\leq0.01\,\mu$Hz).
More recent studies of Vela have not been sensitive to very small glitches \citep[e.g.][who are sensitive only to sizes above $\sim1\,\mu$Hz]{slk+16} or have focussed on resolving large glitches in time with very high cadence observations \citep{fla90,dml02,pdh+18}.

In the following we present a systematic search for small glitches using daily observations of the Vela pulsar. 
We design the search algorithm in a way that it is also sensitive to other types of events, such as changes in spin-down rate only, or other possible combinations of ($\Delta\nu$, $\Delta\dot{\nu}$) signs. 
Hence our results constitute a broad characterisation of the small rotational irregularities seen in the Vela pulsar. 
We present two previously unreported glitches of medium and small size, and detail the population of timing-noise events discovered.

\section{Observations}
\label{data}

For this study, we analysed radio observations of the Vela pulsar that were carried out at the University of Tasmania's Mount Pleasant Radio Observatory (Hobart, Australia) with a 14 metre dish for up to $\sim17$ hours a day between October 1981 and October 2005 \citep{dml02,dlm07}.

\begin{table*}
\caption{Main properties of the three sets of high-cadence TOAs.}
\label{tb:obs}
\centering
\begin{tabular}{lclll}
\hline\hline
Dates                 & MJD range  & Obs. Freqs.     & Cadence    & TOA error     \\
                      &            & (MHz).          & (minutes)  & ($\mu$s)      \\
\hline
Oct. 1981 - May 1986  & 44886 - 46563 & 635          & 2-15       & $7-15$        \\
Jan. 1990 - Sep. 1997 & 47894 - 50705 & 635; 950     & 2; 2       & 30; 22        \\       
Apr. 1999 - Oct. 2005 & 51294 - 53671 & 635; 990     & 2-4; 2     & 30; 25        \\
\hline
\end{tabular}
\tablefoot{The quoted cadences are the most common time lengths between TOAs, which generally correspond to $2$-minute integration observations. 
        Observations were less dense on occasion, however. 
        The quoted TOA errors are the average of all available values when they are (roughly) distributed normally. A range of values is given otherwise.
        }
\end{table*}

\begin{figure}
  \centering
  \includegraphics[width=9cm]{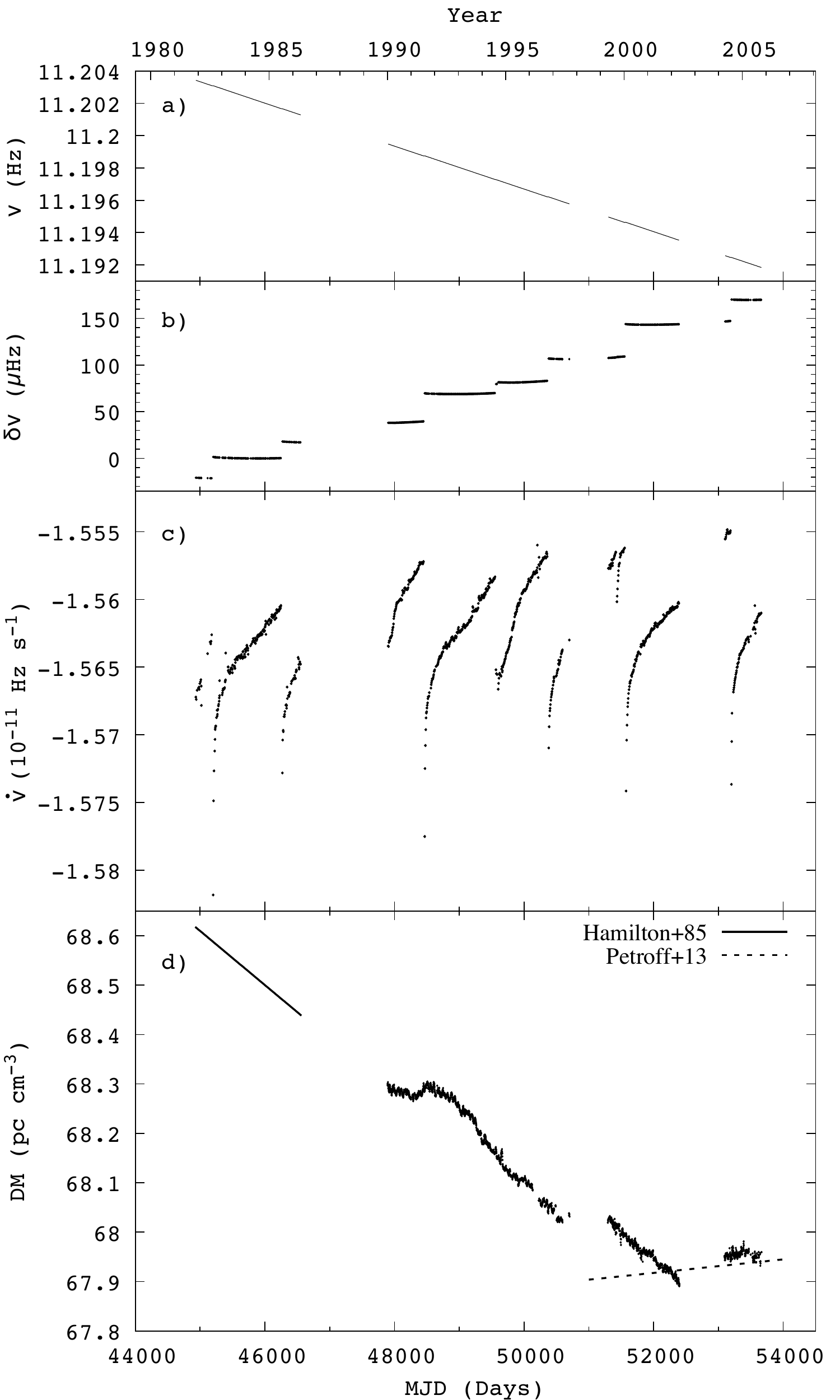}
  \caption{Mount Pleasant observations of the Vela pulsar between 1981 and 2005. 
    \textbf{(a)} The evolution of the spin frequency $\nu$.
    \textbf{(b)} Frequency residuals relative to a linear model for the $700$\,d of data before the second glitch (MJD 46257).
    \textbf{(c)} The evolution of the spin-down rate $\dot{\nu}$.  Both $\nu$ and $\dot{\nu}$ were obtained from fits of a simple slow-down model to TOAs in a moving window covering 20 days, and moving by $5$\,d each stride.
    \textbf{(d)} Daily DM measurements.
    The solid line is the DM model we use for the data obtained in the 1980s. 
    The segmented line is an extrapolation backward from a DM model adjusted to data after MJD 54,000.
  }
  \label{data_p}
\end{figure}

The full dataset spans 24 years, but there are three long periods without data (of $3.6$, $1.6,$ and $1.9$\,yr) that divide the dataset into four sections.
However, based on the observing frequencies and cadence of the observations, we can combine the last two sections and operate over three main datasets.
The three datasets roughly correspond to data taken during the 1980s, 1990s, and 2000s (Table \ref{tb:obs}; Fig.~\ref{data_p}).
When the three longest gaps without data are removed, about 16.9 years of almost continuous observations remain, which include seven large glitches and their recoveries.
The longest uninterrupted period of observations is the dataset from the 1990s, which lasted 7.7 years.

\subsection{Pulse times of arrival}

To track the rotational evolution of the pulsar (by measuring the spin frequency $\nu$ and its time derivatives, a process we hereafter call just `timing'), we used pulse times of arrival (TOAs), each derived from two-minute-long observations \citep[see][for details]{dml02}. 
The observations were designed so that the rotation of the Vela pulsar would be monitored continuously while the source was well above the horizon. 
Consequently, the dataset is further divided into daily blocks of TOAs covering about 0.7 days each, and separated by approximately $7$\,h with no TOAs, with the exception of data from the 1980s, where daily coverage was on average much lower, around $0.1$\,d. 

The highest available density of TOAs is in the dataset from the 1990s, which reaches up to more than 1,000 TOAs per day (combining two observing frequencies, see below). 
A similar density of TOAs is available in the dataset from the 2000s, though observations at $635$\,MHz are sometimes not continuous, but separated by 4-6 minutes.
Additionally, between MJD$\sim$51,300 and $\sim$52,000 (2000's dataset), the observations often cover only $\sim60\%$ of each day ($14.4$\,h).

\subsection{Observing frequencies and dispersion measure}

The availability of TOAs at different observing frequencies changes between datasets.
During the 1980s, only observations at 635\,MHZ were taken, but simultaneous observations at 954\,MHZ started in March 1986 (about two months before the end of the dataset from the 1980s).
During the 1990s, simultaneous observations at 635, 950, and 1390 MHz were carried out over almost the entire interval.
From MJD 49,467 on, the 950\,MHz observations were replaced by 990\,MHZ observations, which continued throughout the dataset from the 2000s (Table \ref{tb:obs}). 
We used only data taken at the two lower frequencies (below $1000$\,MHz), for reasons that we explain in the next section.

Precise daily measurements of the dispersion measure \citep[DM, e.g.][]{lk05} are possible because hundreds of TOAs measured at two different frequencies each day are available. 
The DM values that we obtained agree with the decreasing trend determined by \citet{hhc85}. 
Our results show also that the decreasing trend was momentarily interrupted at least twice by mild increases that lasted 1-2\,yr (Fig.~\ref{data_p}).
The decreasing trend finally ceased by 2002-2003, when a new mild increase is observed.
From then on, the DM seems to have stabilised just below $68$\,pc\,cm$^{-3}$.
This is consistent with some Parkes observations reported by \citet[][see their Fig.~2]{pkj+13}, 
who measured a positive DM slope after MJD 54,000, beyond the end of our data. 
The segmented line in the DM plot in Fig.~\ref{data_p} is an extrapolation backwards from that trend.
We fit for the DM in all our analyses, with the exception of the data from the 1980s, for which we used the DM value and time-derivative measured by \citet{hhc85}.

\section{Pulsar timing}

We used the position and proper motion measured by \citet[][based on very long baseline interferometry (VLBI) observations]{dlrm03}, and fit for DM every time that data at multiple observing frequencies are available.
Phase-connected timing solutions for subsets of data spanning hundreds of days were generated, and their parameters were used as initial values for all the analyses described below. 
The rotational model we used is a standard truncated Taylor series that describes the pulse phase,
\begin{equation}
\phi(t)=\phi_0+\nu(t-t_0)+\frac{\dot{\nu}}{2}(t-t_0)^2+\frac{\ddot{\nu}}{6}(t-t_0)^3 + \cdots \, ,
\label{rotmodel}
\end{equation}
where $t_0$ is a reference time typically chosen at the centre of the subset of data.
Whenever possible, we aimed at generating timing solutions with a constant $\ddot{\nu}$, that is, using only two frequency derivatives.
However, recoveries following large glitches induce rapid changes of the rotational parameters, with $\dot{\nu}$ evolving exponentially for some post-glitch period. 
For the intervals immediately after glitches, we therefore fit three frequency derivatives  (up to $\dddot{\nu}$), which is sufficient to describe the data if subsets of observations are kept short.

\subsection{Data selection}
In rare occasions there are blocks of TOAs that do not phase-align with the surrounding data, but instead require a step in phase $\phi$ to be included.
These phase steps affect both the 650 and 950\,MHz data (or both 650 and 990\,MHz in the 2000s) simultaneously, hence we believe such effects could have been caused by short-term problems with the observatory systems or temporary changes of the fiducial point of the pulse templates, for instance.  
In most cases it was possible to align the misaligned data by fitting for a phase jump just before the start of a particular block of data together with a phase jump of opposite sign right at its end. 
The TOAs for which this was not straightforward to do were removed from further analyses.

We avoided using the plethora of $1390$\,MHz observations available in the datasets from the 1990s and 2000s for the timing analyses because they are not phase aligned with the lower frequency observations. 
This misalignment is not solved by fitting for DM and would require modifying the phases of the 1390\, MHz TOAs. 
The misalignment is not constant with time. Instead, the 1390\,MHz TOAs are seen to phase drift with respect to the lower frequency TOAs over the course of each day. This effect cannot be simply accounted for by a constant phase jump.  
Such drifts might have been produced by the intrinsic evolution of the linear polarisation angle across the phase of the pulse, in combination with the fact that the receiver could only detect one linear polarisation and the feed rotates with respect to the source during the day.
Correcting for this effect in data taken decades ago is complex, therefore we neglected observations at 1390\, MHz and worked only with TOAs measured at 635 and 950 (or 990) MHz, which are free from such misalignments.
Furthermore, the daily root mean square (RMS) of the phase residuals relative to a simple slow-down timing model (Eq. \ref{rotmodel}) is larger at 1390\,MHz because the signal-to-noise ratio is lower, which might degrade the precision of the glitch searches.

\subsection{Daily effective times of arrival}
\label{sc:tzr}
For the type of analysis we wished to perform, we chose to reduce the 635 and 990-950\,MHz TOAs to a single-TOA-per-day dataset.
There are multiple benefits of doing so: it averages out any undetected daily phase drifts such as those affecting the 1390\,MHz data (see above); the generated dataset has a rather constant TOA cadence, which facilitates determining glitch detectability limits, and greatly reduces the computing time during the searches. 
We note that due to the relatively large RMS of the high-cadence TOAs (with respect to a simple model, see below), the sensitivity of the glitch search is not negatively affected by the use of the less dense daily effective TOAs. This process restricts the timescales we can probe in the timing analysis, but minuscule, short-lived ($\leq3$\,d) transient irregularities are beyond the scope of this study.

For better precision, we decided to use TOAs generated only for the days that are well covered by the observations. 
To this end, only daily blocks of minimum 850 TOAs (which translates into over 60\% daily coverage) were used to produce this secondary dataset. 
This limit works well for the dataset from the 1990s, but the data cadence in the 2000s is lower. To include blocks with the same minimum daily coverage, we therefore reduced the minimum to 300 TOAs per day.
For the dataset from the 1980s, less restrictive conditions were used, and we required a minimum of 4 TOAs. 
These conditions automatically separate blocks of $\sim17\,$h, thereby excluding cases in which the observations do not cover a good fraction of the day or are not dense enough.

Each selected TOA block was fitted to a timing model (Eq. \ref{rotmodel}) with one frequency derivative, using the timing package {\sc tempo2} \citep{hem06,ehm06}. 
The effects of a varying $\dot{\nu}$ over this short time ($<1\,$d) can be safely neglected, therefore we omitted the second frequency derivative from the fit. 
Typical weighted RMS$_w$\footnote{The weighted RMS of a collection of $N$ values $x_i$ with errors $\delta x_i$ is calculated as $\textrm{RMS}_w = \sqrt{\sum_i^N (x_i/\delta x_i)^2 / \sum_i^N \delta x_i^{-2}}$ }
values of these fits are in the range $50$-$70\,\mu$s for the data from the 1990s and 2000s, but they can reach up to 150\,$\mu$s for the data from the 1980s (attributable to a less stable observatory clock used in those days).
These dispersions are larger than the typical TOA uncertainty (of $20$-$30\,\mu$s, see Table \ref{tb:obs}).
We note that our RMS$_w$ values are consistent with the daily fits reported by \citet[][]{pde+16}, who obtained daily RMS values of about 50\,$\mu$s (they used a larger telescope and arguably a more sensitive observing system). 
Single-pulse timing of the Vela pulsar offers phase residuals with larger RMS of about $0.3$\,ms over 0.5\,h \citep{pdh+18}.
The timing precision is increased when very short segments of data are combined, which averages out the effects of single-pulse shape variations and jitter.

The standard output of {\sc tempo2} includes the parameters TZRMJD, TZRFRQ, and TZRSITE, which define a TOA that has zero residuals relative to the rotational model adjusted at the time (regardless of what parameters were varied). 
While the last two parameters are the observing frequency and site, respectively, the former (TZRMJD) is an \emph{\textup{ideal}} pulse time of arrival, as determined by the best-fit timing model.
It is calculated as the time of arrival of the first TOA after the centre of the fitted time interval, but with its residual set to zero (i.e. `corrected' according to the best-fit model). 
The TZRMJD values of the selected blocks of TOAs are used as our effective TOAs and constitute the secondary dataset used for the glitch search described below.

Error bars for the TZR TOAs were defined as the RMS$_w$ of the timing residuals of the block of TOAs that was used to define TZRMJD.
Several collections of residuals corresponding to daily blocks were inspected, and it was verified that the residuals are distributed normally. 
The RMS$_w$ values are generally about two to three times higher than the typical 2-minute TOA error bars.
It is possible that the assigned error bars overestimate the TZR TOA uncertainty.
However, we kept the use of RMS$_w$ because the effect of these error bars on the results of this investigation is essentially negligible.

\section{Method}
\label{methods}

To search for possible rotational irregularities, we followed a procedure similar to the one presented in \citet{eas+14}.
We started by assuming that a change in rotational parameters $\nu$ and/or $\dot{\nu}$ occurred after every single TOA. Then, a simple model reflecting this change was fitted to the data.
The model has only two parameters: a step change in frequency $\Delta\nu,$ and a step change in spin-down rate $\Delta\dot{\nu}$. 
If the fit to the glitch model is considered acceptable, then an irregularity may have occurred, and the event is flagged as a candidate.

The algorithm works as follows: First, 20 days of data are selected and a fit for a rotational model like Eq. \ref{rotmodel} is performed using {\sc tempo2}.
We call this set of data the \emph{\textup{first}} set.
The $\ddot{\nu}$ term in Eq. \ref{rotmodel} is included because in $20$\,d the contribution of this term to the phase residuals can be significant: roughly between $9$ and $90$\,ms for $\ddot{\nu}\sim0.1-1\times10^{-21}$\,Hz\,s$^{-2}$ \citep[which is a typical value between glitches; e.g.][]{cdk88,els17}. 
Next, a \emph{\textup{second}} set of data is defined with all the TOAs in a window of $N_d$ days starting at the first TOA after the end of the first set.
We tried both $N_d=10$ and $20$\,d. The residuals of these TOAs relative to the best-fit model for the first set are calculated. 
If a glitch occurred between the two sets, then the residuals of the second set should behave according to 
\begin{equation}
\Delta\phi_\textrm{g}(t)=-\Delta\nu (t-t_\textrm{g}) -\Delta\dot{\nu}\frac{(t-t_\textrm{g})^2}{2} ,
         \, \quad(t>t_\textrm{g})\,,
\label{glitmod}
\end{equation}
where $t_\textrm{g}$ is the time of the glitch, which we take as the last TOA of the first set.
We fit this function to the residuals of the second set, and in addition, we also separately fit a linear version, with $\Delta\dot{\nu}=0$, as well as a version with $\Delta\nu=0$.
The fit that returns the smallest reduced $\chi^2$ is selected as the best representation of the second set. 
When the best model has been established, further checks are performed to assess whether a timing irregularity might be present between the two sets, and the candidate event is either stored or discarded. The search then continues by moving the start of the first set one TOA forward, defining the new datasets, and repeating the process.

\begin{figure}
\centering
\includegraphics[width=9.2cm]{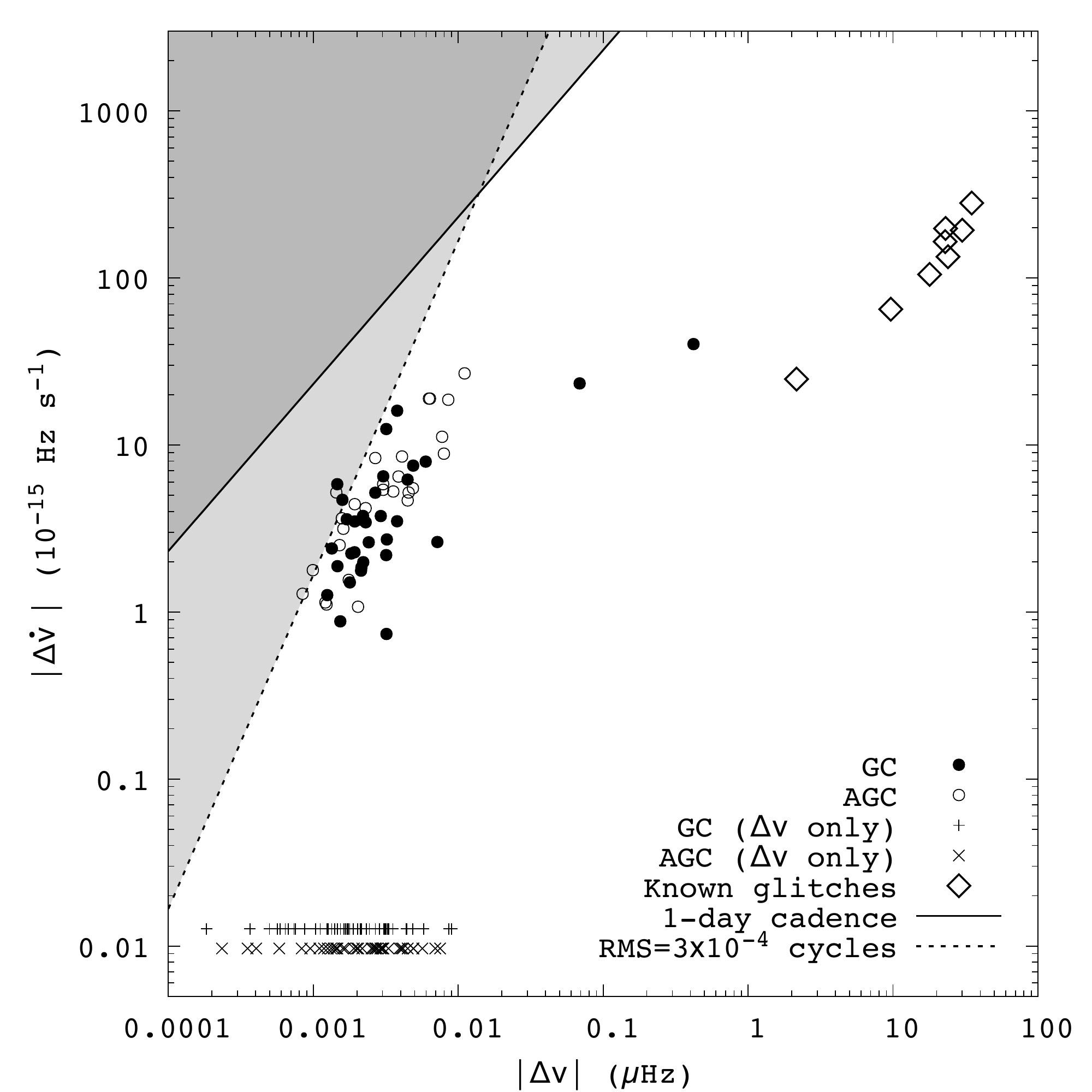}
\caption{Selected GCs and AGCs for $N_d=10$\,d. 
Candidates without detected $\Delta\dot{\nu}$ are plotted with crosses and the letter {\sf x} using artificially assigned fixed values of $\Delta\dot{\nu}$ to place them at the bottom of the plot. 
Previously known Vela glitches that are in the dataset we used are plotted with diamonds. 
Straight lines indicate detection limits that depend upon cadence and noise. 
They  indicate that detections above them are not expected.
Error bars are omitted for clarity. 
The two new glitches we found are easily recognisable. 
They stand out as the GCs with the highest $\Delta\nu$ values. }
\label{df0df1}
\end{figure}


To decide whether a significant change in rotational parameters occurred, thus a candidate event should be flagged, certain criteria should be met. First, phase continuity must be preserved between the solutions for the first and second set\footnote{We note here that in general, and depending on TOA coverage, phase continuity can be lost when a large glitch occurs; such large events are easily identified in the residuals and are not the target of our search.}. 
We required the phases $\phi(t)$ returned by each model to nearly meet within a certain time range that we defined for our purposes as the interval between the last TOA of the first set and first TOA of the second set. 
We did not impose an exact matching of the predicted phases in order to account for the uncertainty of the event epoch, and because we should allow for some low-level noise as well. 
The maximum allowed difference in phase was taken to be $2\,\sigma_\phi$, where $\sigma_\phi$ is the RMS$_w$ of the phase residuals of the first set.

Secondly, the examined subsets should not be separated by a long period without any TOAs. 
We set the maximum allowed separation at $6$ days. 
This cutoff is by no means unique and was decided empirically, but varying it within a reasonable range has negligible effect on the results for this particular (dense) dataset that we examine here.

Cases in which the amplitude at the extremum of the quadratic function (Eq. \ref{glitmod}), or equivalently, the amplitude of the linear function after $N_d$ days, was lower than $2.5\sigma_\phi$ were not considered.
Moreover, when the selected second model was quadratic, its extremum should not occur at a time before the start of the second set.

The quality of the fits was also used to decide whether the candidate event should be flagged. 
To assess the fits, we used the RMS$_w$ of both sets of data. 
RMS$_w$ values are preferred over $\chi^2$ because the error bars of the TZR TOAs could be overestimated (see above), which could produce particularly low $\chi^2$ values that would not serve as an indicator for goodness of fit (almost all fits give $\chi^2$ values lower than $0.5$ and distributed around $0.1$).  
About 80\% of the fits for the first set of data have residuals with dispersions RMS$_{w1}$ in the range 
$0.04\times10^{-3}\leq \textrm{RMS}_{w1} \leq0.2\times10^{-3}$ cycles, 10\% have $0.2\times10^{-3}\leq \textrm{RMS}_{w1} \leq0.3\times10^{-3}$ cycles, and the remaining 10\% have 
$0.3\times10^{-3}\leq \textrm{RMS}_{w1} \leq 1.0\times10^{-3}$ cycles.
With the aim of being somewhat permissive in terms of accepting candidates, we chose to set a maximum residuals dispersion of 
$\sigma_\phi^\textrm{max}=0.3\times10^{-3}$ cycles ($\sim0.03$\,ms in Vela) to separate acceptable from unacceptable fits.
For the second set, the residuals present a similar distribution of RMS$_w$ values.

The final restriction to select candidates was that the fit to the second set must give residuals that satisfy $\textrm{RMS}_{w2} \leq 3\times\textrm{RMS}_{w1}$.
This last criterion makes assessment of a candidate dependent on the level of noise present in the first set.
We expect that if the first fit is particularly good or noisy, then the second fit should be similar.
The factor of $3.0$ is to relax this restriction and accept as candidates even those events in which the behaviour perhaps only remotely approaches that of a glitch or the selected model. 
For example, the condition allows the detection of relatively large glitches that may be followed by a small exponential recovery that might cause the data to deviate from the simple model in Eq. \ref{glitmod}.

\section{Results}

To begin with, we focused on events with either a glitch-like signature, with $\Delta\nu>0$ and $\Delta\dot{\nu}\leq0$, or \emph{\textup{anti-glitch}} like, which we defined as $\Delta\nu<0$ and $\Delta\dot{\nu}\geq0$. 
This includes events with $\Delta\dot{\nu}=0$.
Other types of features, including changes in spin-down rate alone, are discussed in the next section.

Following the procedure outlined in section \ref{methods}, we first selected $N_d=10$\,d as the length of the second set. 
This resulted in the selection of 151 glitch candidates (GCs) and 159 anti-glitch candidates (AGCs). 
Because the process moves at just one TOA at a time, these events include groups of candidates that are very close in time, which could correspond to multiple detections of a single event \citep{eas+14}. 
Conversely, some of these cases might correspond to two or more independent events that occurred in quick succession. 
Unfortunately, it is not always possible to distinguish between the two scenarios, and thus we proceeded with the assumption that all candidates whose $t_g$ times differ by three or fewer days correspond to the same event. 
Of these, we only kept the candidate that offered the best fit to the data, that is, the solution that leads to the smallest residual dispersion measured as $\sqrt{\textrm{RMS}_{w1}^2+\textrm{RMS}_{w2}^2}$.
In most cases, potential multiple detections consist of only a pair of candidates, whilst the maximum numbers of events grouped together was $9$, which covered nearly $10$\,d in total. 
When the duplicates were removed, 83 GCs and 66 AGCs remained.

Events with an undetectable $\Delta\dot{\nu}$ represent an important fraction: 51 GCs and 38 AGCs. That is, 61\% and 58\% of the candidates, respectively. 
The $\Delta\nu$ and $\Delta\dot{\nu}$ measured steps for all candidates are plotted in Fig.~\ref{df0df1}, where for plotting purposes we assigned an artificial $|\Delta\dot{\nu}|$ value of about $0.01\times10^{-15}$\,Hz\,s$^{-1}$ for the candidates in which this quantity could not be measured.

We repeated the analysis for $N_d=20$\,d.
This led to fewer candidates: 50 GCs and 45 AGCs (after multiple detections were removed).  
One possible cause for this is that the longer time of the second set implies higher chances that noise or new glitch-like events affect the data (see the above selection criteria).  
In this case, however, the fraction of candidates with undetected $\Delta\dot{\nu}$ steps is smaller: 30\% of all GCs and 33\% of all AGCs.
This is likely explained by the fact that longer second sets allow a better determination of $\Delta\dot{\nu}$.
These candidates distribute in the $\Delta\nu$-$\Delta\dot{\nu}$ space in the the same way as the candidates for $N_d=10$\,d.

For completion, in Fig. \ref{df0df1} we also include the large glitches (already known) that occurred in the time intervals for which we have data\footnote{We did not have data for the large glitch at MJD\,$\sim47519.8$, hence it is not plotted.}.
All of them have $\Delta\nu>1\,\mu$Hz and are followed by exponential-like recoveries. 
Because of their size and strong exponential signature, we did not attempt to redetect the largest glitches with the detector. 
For consistency with the other plotted data, however, we display glitch parameter magnitudes measured with the same method as for the GCs, that is, by using $20$\,d to calculate a base rotational model and then modelling the phase residuals of the following $N_d$ days of data with Eq. \ref{glitmod}.
The only difference is that for the known glitches we used $N_d=20$\,d, because for the large glitch near MJD 46257 there are insufficient TOAs during the first $10$\,d.
The same measurements are also used in Figs. \ref{histos} and \ref{fg:rr4}. 
Therefore we note that the glitch parameters presented in Fig.~\ref{df0df1} are not optimally calculated and should not be used for theoretical glitch studies; nonetheless, the differences of our $\Delta\nu$ values with the published results are lower than $4$\%.

\subsection{Search sensitivity}

Variations in the search algorithm or the criteria (section \ref{methods}) used to select candidates do not produce substantially different results. 
By this we mean that qualitatively, the results remain the same, with very similar distributions in magnitudes. Nonetheless, for each individual event, small changes in the measurement procedure such as the inclusion of a third frequency derivative in the timing model or a different length of the fitted intervals would produce slightly different results.

We see no obvious trends in terms of the distribution of candidates in the $\Delta\dot{\nu}$--$\Delta\nu$ space when $\sigma_\phi^\textrm{max}$ is made smaller. 
Similarly, when the factor $3.0$ in the condition $RMS_{w2}<3\times RMS_{w1}$ is reduced, the number of candidates decreases, as expected, but without a clear general pattern.
However, one important effect is that half of the previously known (large) glitches in the analysed data would not be selected if this factor were smaller than $3.0$.
This is likely because larger events involve exponential-like recoveries and step-like changes $\Delta\ddot{\nu}$, which are not included in the model, thereby making the residuals larger.
On the other hand, if the factor $3.0$ is made very large, the increase in the number of candidates is marginal, 
the distribution of $|\Delta\nu|$ values for all candidates remains very similar, and there are no additional candidates with sizes larger than $0.02\,\mu$Hz.

The indicative detection limits defined by the straight lines in Fig.~\ref{df0df1} follow the formulae in \citet{eas+14}. Events with parameters in the areas above them cannot typically be detected. 
One of the lines is determined by the cadence of the observations, and we used $1$\,d as a representative number. 
The other limit depends on the accuracy of the rotational model calculated for the first set of data. 
To select valid candidates, the condition RMS$_{w1}<\sigma_\phi^\textrm{max}$ was used (see above), hence we used this value to illustrate our detection capabilities.
The top left and top middle panels in Fig.~\ref{res3}  show some examples of the GCs and AGCs with the lowest $\Delta\nu$ values, which are close to the limit of what the automated method detects as significant.
The main effect on detectability is given by the dispersion of the timing residuals and not by the one-day cadence. 
Thus, while it is highly possible that a large number of events above and to the left of the detection lines were not discovered, the strong reduction in number of candidates with sizes above $\sim0.01\,\mu$Hz seems to be intrinsic to the pulsar.
Examples of phase residuals for GCs and a AGCs are presented in Fig.~\ref{res3} (see also Figs. \ref{fg:glit1res} and \ref{fg:glit2res}). 
Features with $\Delta\nu\sim0.01\,\mu$Hz or larger clearly leave signatures stronger than the typical noise. 
Hence, if there were more of these events, they would have been detected either by eye or by the automated method, or by both (as is the case for the two largest GCs, see below). 
We infer that there is little activity above $\Delta\nu\sim0.01\,\mu${\rm Hz} and below $\sim10\,\mu${\rm Hz} in the form of anti-glitches and noise-like features (no events detected), whilst less than one-third of the known Vela glitches have $\Delta\nu$ in this range.

\subsection{Size distributions of GCs and AGCs}

\begin{figure}
  \centering
  \includegraphics[width=9cm]{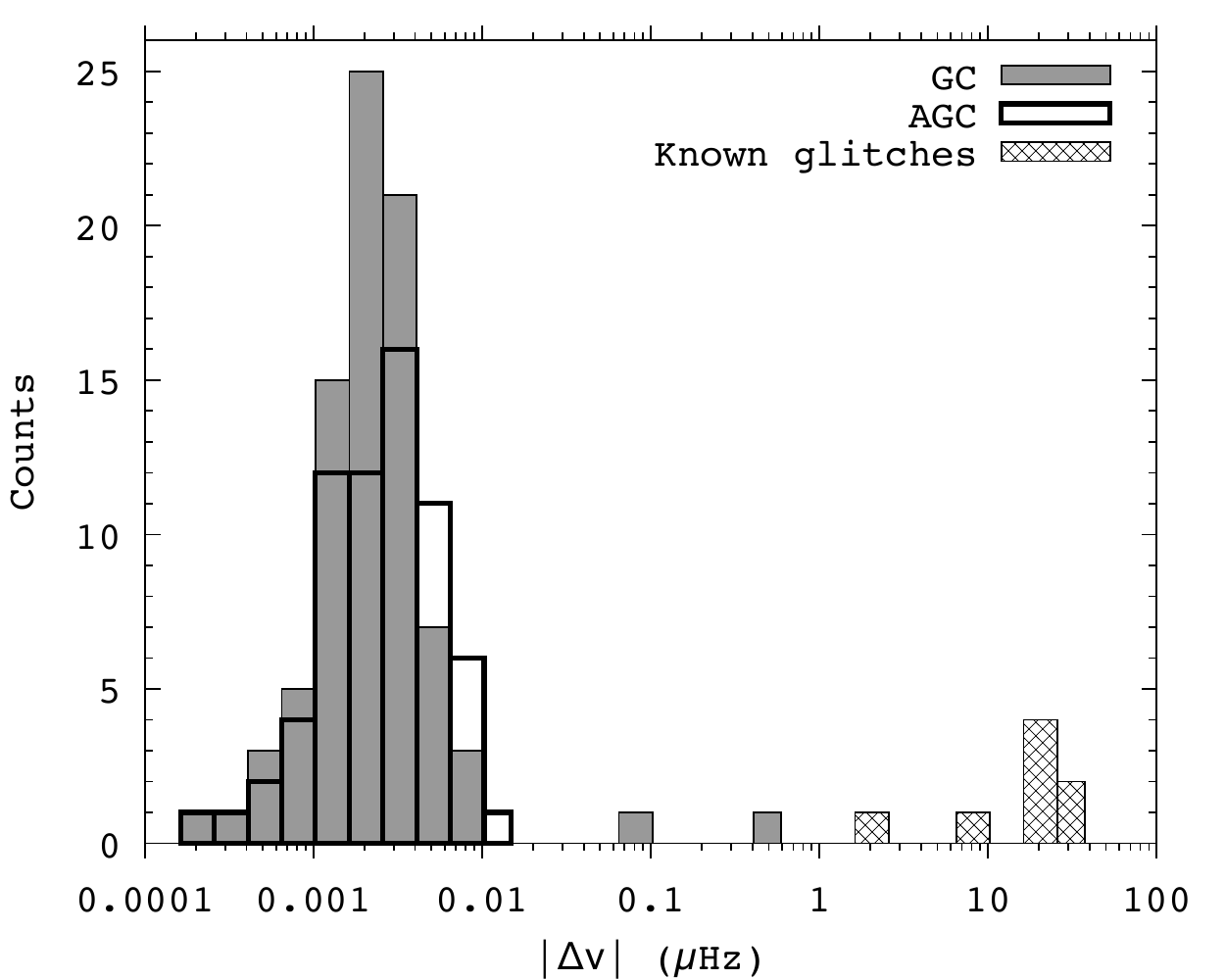}
  \caption{Histogram of $|\Delta\nu|$ for GCs and AGCs, including the eight previously known Vela glitches that are in our data (measured as for Fig. \ref{df0df1}).
}
  \label{histos}
\end{figure}

Histograms of GCs and AGCs sizes are presented in Fig.~\ref{histos}. 
We tried to fit both size distributions with four different models: log-normal, Weibull, Gaussian, and power-law distributions.
The log-normal distribution is favoured among them, for which a Kolmogorov-Smirnov (KS) test returns $p_\textrm{KS}$ values $0.02$ and $0.98$ for GCs and AGCs, respectively.
The low $p_\textrm{KS}$ for GCs arises because of the two largest candidates, which are clearly seen in the histogram  as outliers at sizes $\Delta\nu\sim0.07$ and $\sim0.4\,\mu$Hz. 
When this pair is excluded, then $p_\textrm{KS}$ for a log-normal distribution rises to $0.44$. 
It is worth mentioning that the true shape of the lower end of the distributions is poorly constrained, therefore any modelling of GCs and AGCs distributions is only approximate.

As discussed in the following, the largest of the GCs is most likely a typical glitch. 
This is  supported by its large size (closer to those of the known glitches) and by the fact that it is followed by an exponential recovery. 
This new glitch has not been reported before, and we present more details about it below.
On the other hand, the nature of the second largest GC is less clear. It could be either a very small glitch or a particularly large noise event, although its signature is characteristically glitch-like. 
Using the best-fit log-normal distribution for GCs (excluding the two largest events), we can calculate the probability for events of its size ($\Delta\nu=0.0621\pm0.0003\,\mu$Hz, Table \ref{tb:glit1}), which is found to be $10^{-8}$. This favours the interpretation of this event as a typical glitch, which does not belong to the same population as the other GCs. 

A KS test of the $|\Delta\nu|$ values for GCs and AGCs indicates a probability of only $27.6$\% that both samples were drawn from the same distribution.
The probability is further reduced to $16.5$\% when the two largest GCs are removed.
For the $|\Delta\dot{\nu}|$ values the KS probability is $\leq7\times10^{-5}$ in both cases.

\subsection{Previously unreported glitches}

The detector found only two events with $\Delta\nu>0.02\,\mu$Hz.
The same two events immediately stood out during visual inspections of the timing residuals.
Based on the analysis above and the properties described below, we conclude that these are standard glitches that have not been reported before. 
To properly describe them, we performed additional analyses around their epoch. 

The TOAs were modelled with a rotational model that is a combination of the function defined in Eq. \ref{rotmodel} for the TOAs prior to the glitch, and a glitch model for the data after the glitch,
\begin{equation}
  \Phi(t)=\phi(t) + \Phi_\textrm{g}(t)H(t-t_\textrm{g}) \quad,
  \label{eqglit}
\end{equation}
where $H$ is the Heaviside step function and 
\begin{eqnarray}
  \Phi_g(t)&=&\Delta\phi + \Delta\nu_\textrm{p} (t-t_\textrm{g}) + \Delta\dot{\nu}_\textrm{p}\frac{(t-t_\textrm{g})^2}{2} + \Delta\ddot{\nu}_\textrm{p}\frac{(t-t_\textrm{g})^3}{6}  \nonumber \\
           &&-\tau_\textrm{d}\Delta\nu_\textrm{d}e^{-(t-t_\textrm{g})/\tau_\textrm{d}} \quad.
  \label{eqglit2}
\end{eqnarray}

This glitch model includes \emph{\textup{persistent}} steps in the spin frequency and its first two derivatives, as well as an exponential post-glitch recovery.
The changes at the glitch epoch are $\Delta\nu=\Delta\nu_\textrm{p}+\Delta\nu_\textrm{d}$, $\Delta\dot{\nu}=\Delta\dot{\nu}_\textrm{p}-\Delta\nu_\textrm{d}/\tau_\textrm{d}$, and $\Delta\ddot{\nu}=\Delta\ddot{\nu}_\textrm{p}+\Delta\nu_\textrm{d}/\tau_\textrm{d}^2$.
The parameter $\Delta\phi$ is a phase step at the glitch that accounts for our ignorance of the exact glitch epoch $t_\textrm{g}$, which was initially set at the centre of the gap between the pre-glitch and post-glitch TOAs.
The third frequency derivative was set to zero in these analyses.

\subsubsection{New very small glitch in 1991}

The presence of this glitch is hard to miss because of its clear, sharp feature, which is indicative of a discontinuity in the spin rate that stands out in the phase residuals for over $200$\,d around its epoch.
It occurred near MJD 48550, that is, 93 days after the large glitch in 1991.

\begin{figure}
\centering
\includegraphics[width=9cm]{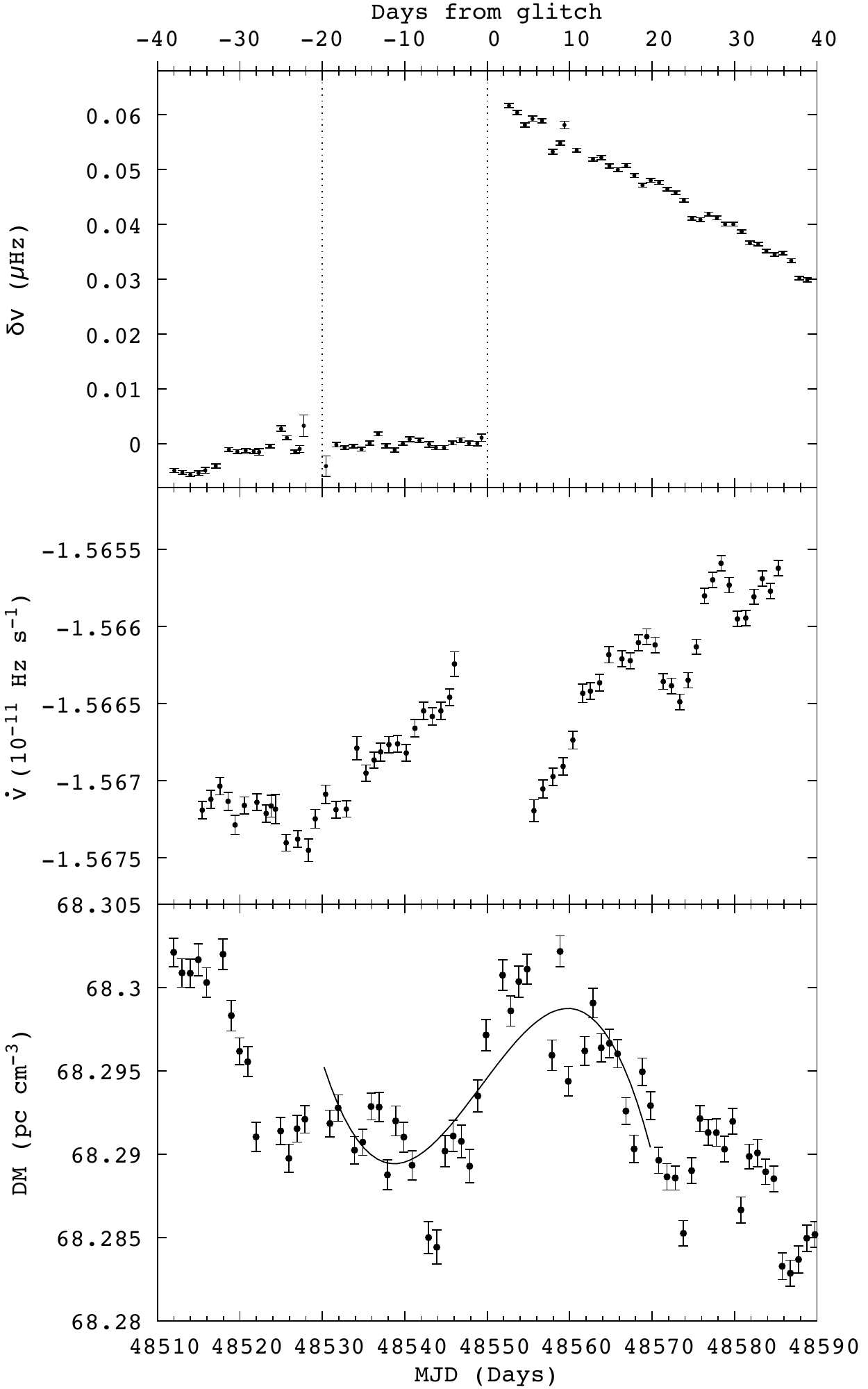}
\caption{Rotational behaviour and DM evolution over $80$\,d around the new small glitch at MJD $48550.37$.
{\sl Top}: Frequency residuals relative to a quadratic model adjusted to $\nu$ data from $20$\,d before and up to the glitch epoch (interval marked by vertical dotted lines).
The $\nu$ values come from fits to TOAs in a moving window $3$\,d long that moves by $1$\,d each stride. 
{\sl Middle}: Spin-down rate $\dot{\nu}$.
Values come from fits to TOAs in a moving window of length $10$\,d and that moves $1$\,d with each stride.
Windows that contained the glitch epoch were not considered, hence the gap in the plots.
{\sl Bottom}: Daily values of the dispersion measure and the model we used to describe its variation for the $40$\,d around the glitch. 
}
\label{fg:glit1}
\end{figure}

To characterise the event, we gathered all TOAs in a 20\,d radius around MJD 48550, and fit $\Phi$ to the data.  
We preliminarily set $\Delta\ddot{\nu}_\textrm{p}=0$, $\Delta\nu_\textrm{d}=0$. 
All the varied parameters are detected well and are consistent with the output of the automatic search. 
Then we considered other factors that could affect this measurement and then determined the glitch epoch more precisely.

\begin{figure}
\centering
\includegraphics[width=9cm]{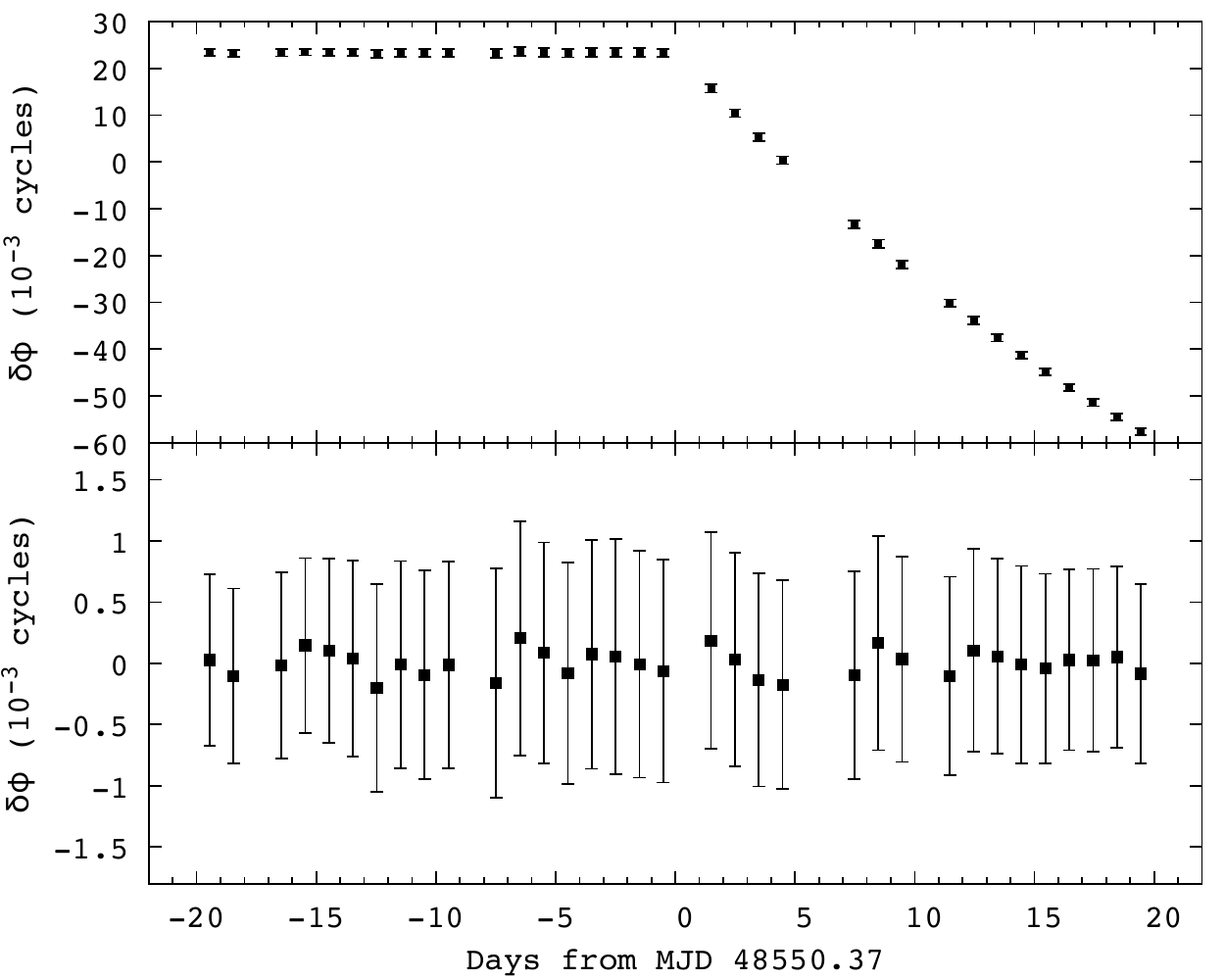}
\caption{Phase residuals 40\,d around the new small glitch at MJD $48550.37$.
{\sl Top}: Residuals relative to a model fitted to the 20\,d of data prior to the glitch.
{\sl Bottom}: Residuals relative to the glitch model in Table \ref{tb:glit1}.}
\label{fg:glit1res}
\end{figure}

The DM around the time of the glitch behaves erratically (Fig.~\ref{data_p}), and in particular, it increases rapidly by nearly $0.02$\,pc\,cm$^{-3}$ through the glitch (Fig.~\ref{fg:glit1}).
Several changes like this can be observed in a 300\,d window centred on the glitch, hence we reject a connection between the DM change and the glitch itself.
In order to reduce the (small) effects that the DM variations could have on the glitch measurement, we used {\sc tempo2} to model the DM evolution in a $40$\,d window centred on the glitch.
For this we used the high-cadence original TOAs at 635 and 950\,MHz because they can offer a precise DM determination. 
The rotational model was the same as above, and the DM model was a third-order polynomial around $t_0$ (Fig.~\ref{fg:glit1}), whose coefficients can be found in Table \ref{tb:glit1}.
We note that while the model fails to precisely describe the behaviour, it offers residuals that are small enough ($<0.005$\,pc\,cm$^{-3}$) to not affect the glitch measurement significantly. 
The DM model was then kept fixed during the fits described below.

To determine the glitch epoch, we performed multiple fits for $\Delta\phi$, $\Delta\nu_\textrm{p}$, and $\Delta\dot{\nu}_\textrm{p}$, varying the glitch epoch between the TOAs before and after the glitch, and using a short time step ($10^{-3}$\,d).
The glitch epoch was chosen as the time when $\Delta\phi$ is minimised, and its uncertainty was taken as the range of time over which the error of $\Delta\phi$ is larger than its value.
The results of a final fit to the daily TOAs of all parameters (with the exception of $\Delta\phi$, which was now set to 0.0), using this glitch epoch and the DM model determined above, are listed in Table \ref{tb:glit1} and the post-fit phase residuals in Fig.~\ref{fg:glit1res}.

We searched for a possible change in $\ddot{\nu}$ at the glitch, but found no strong evidence for it.
When 40 rather than 20\,d around the glitch are used, it is possible to marginally measure $\Delta\ddot{\nu}$. 
However,  this might be due to a background of $\dot{\nu}$ fluctuations that populate the data at almost all times (e.g. middle panel in Fig.~\ref{fg:glit1}; see also Fig.~\ref{fg:rot1}), or due to the underlying evolution of $\ddot{\nu}$ caused by the relaxation of the large glitch 93 days before.
Thus, it is not possible to attribute the change to the glitch with certainty. 
There is no clear evidence for an exponential-like recovery (i.e. $\Delta\nu_\textrm{d}=0$) either.

\begin{table}
\caption{Parameters of the new small 1991 glitch.}
\label{tb:glit1}
\centering
\begin{tabular}{ll}
\hline\hline
Parameter                                             & Value      \\
\hline
Time range (MJD)    &  48530.92-48569.81\\
Epoch $t_0$ (MJD)    &  48540.00  \\
$\nu$ (Hz)   &  11.19865161705(5) \\
$\dot{\nu}$ ($10^{-15}$\,Hz\,s$^{-1}$)    &  -15666.1(2) \\
$\ddot{\nu}$ ($10^{-21}$\,Hz\,s$^{-2}$)    &  7.7(1) \\
DM$_0$ (pc\,cm$^{-3}$)   &  68.2896(3) \\
DM$_1$ (pc\,cm$^{-3}$\,yr$^{-1}$)   &  0.06(1) \\
DM$_2$ (pc\,cm$^{-3}$\,yr$^{-2}$)   &  7.3(5) \\
DM$_3$ (pc\,cm$^{-3}$\,yr$^{-3}$)   &  -97(6) \\
Glitch epoch (MJD)     &  48550.37(2) \\
$\Delta\nu _\textrm{p}$ ($\mu$Hz)   &  0.0621(3) \\
$\Delta\dot{\nu}_\textrm{p}$ ($10^{-15}$\,Hz\,s$^{-1}$)   &  -16.5(2) \\
RMS$_w$ ($\mu$s)     &  8.7 \\
Number of TOAs     &  34 \\
\hline
\end{tabular}
\tablefoot{1$\sigma$ error bars are in parentheses and represent the uncertainties of the last quoted digits.
DM$_i$ is the $ith^{\,\rm{}}$ coefficient of the DM model $\rm{DM}(t)=\sum_{i=0}^3 \rm{DM}_i(t-t_0)^i$.
}
\end{table}

\subsubsection{New small glitch in 1999 with detectable relaxation}

\begin{figure}
\centering
\includegraphics[width=9cm]{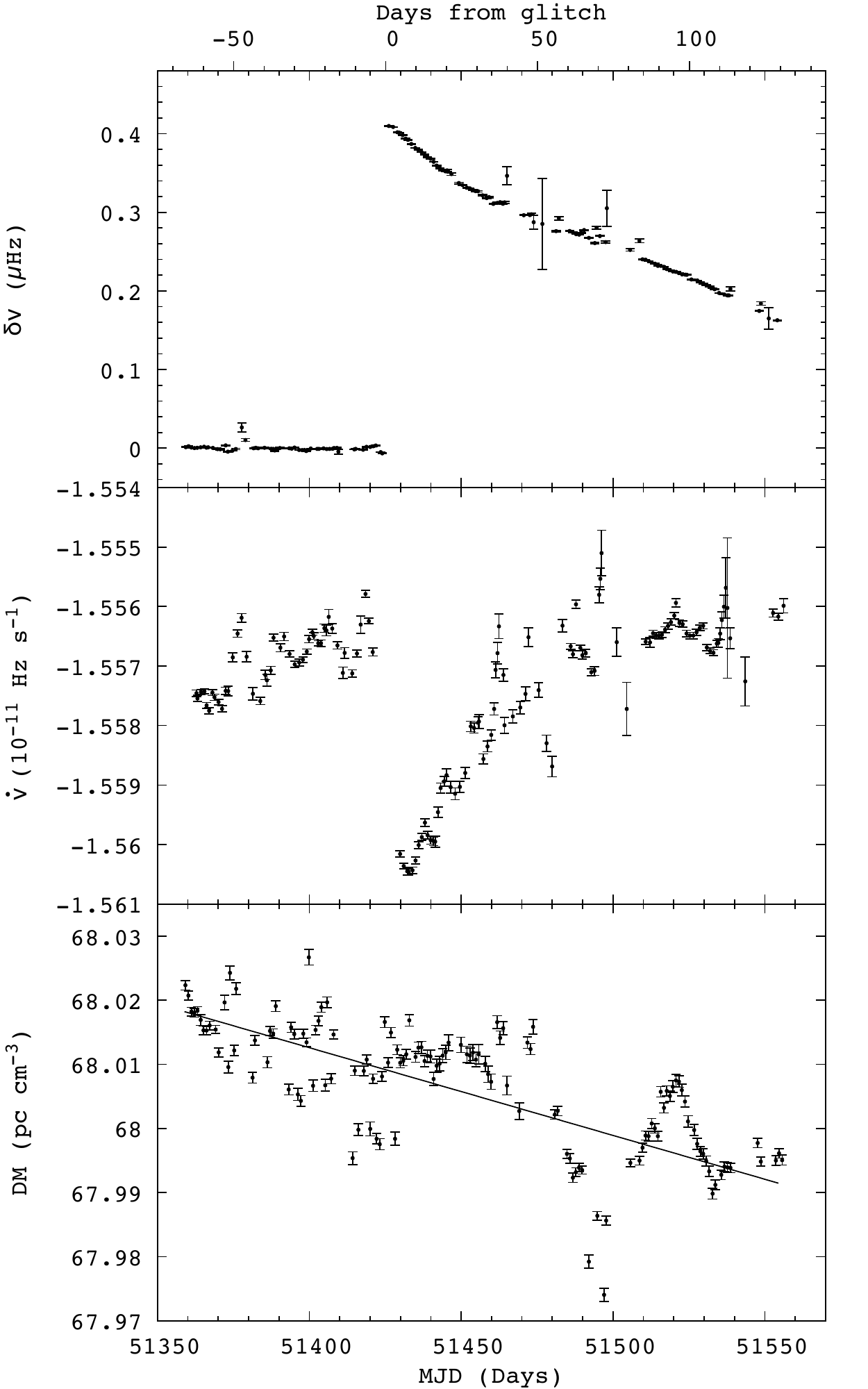}
\caption{Rotational behaviour and DM evolution around the new glitch at MJD $51425.12$.
{\sl Top}: Frequency residuals relative to a quadratic model fitted to $\nu$ data before the glitch. 
{\sl Middle}: Spin-down rate $\dot{\nu}$. 
Both $\nu$ and $\dot{\nu}$ values were measured as in Fig.~\ref{fg:glit1}.
Large error bars are normally produced by gaps in the set of TOAs.
{\sl Bottom}: Dispersion measure and the model used to describe its variation around the glitch. 
}
\label{fg:glit2}
\end{figure}

\begin{figure}
\centering
\includegraphics[width=9cm]{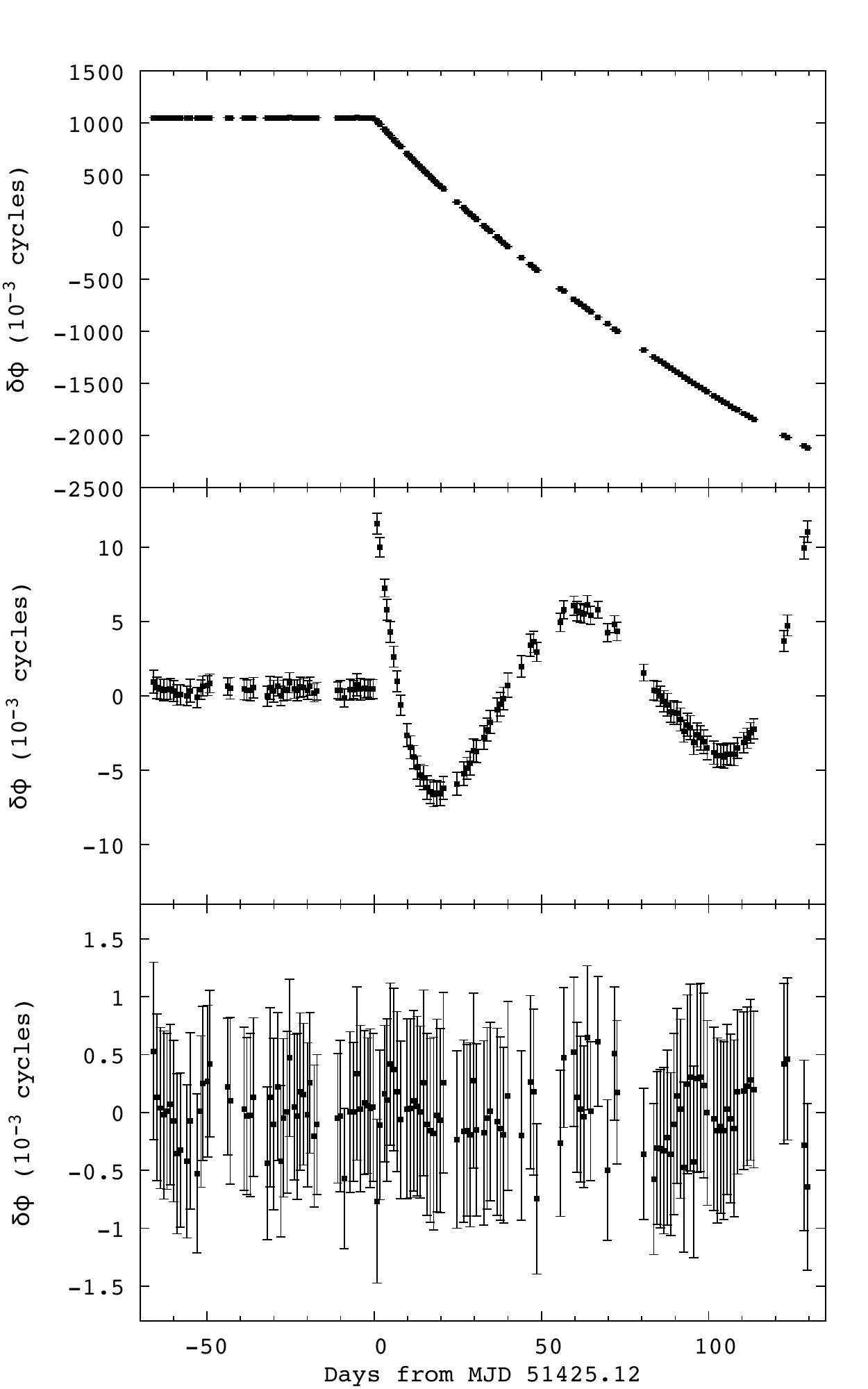}
\caption{Phase residuals around the new glitch at MJD $51425.12$. 
{\sl Top}: Residuals relative to the model in Eq. \ref{rotmodel} fitted to the first 67\,d of data, before the glitch.
{\sl Middle}: Residuals relative to a glitch model without a recovery. 
{\sl Bottom}: Residuals relative to a model that includes an exponential recovery, as in Table \ref{tb:glit2}.}
\label{fg:glit2res}
\end{figure}

This glitch occurred near MJD 51425, that is, 134 days before the large glitch in 2000.
As a first step, we used $40$\,d of data centred on this epoch to fit only for $\Delta\phi$, $\Delta\nu _\textrm{p}$, and $\Delta\dot{\nu}_\textrm{p}$. 
The results obtained are consistent with the output of the automatic search.
An interesting property of this glitch is that there is an exponential-like recovery of the rotation after the event.
This recovery is seen clearly in the spin-down rate evolution shown in Figs. \ref{data_p} and \ref{fg:glit2} and in the middle panel of Fig.~\ref{fg:glit2res}, where large phase residuals after the glitch remain when the recovery is not included in the model.
We modelled this recovery with a single exponential function (Eq. \ref{eqglit2}).

To fit for the post-glitch relaxation, a longer dataset is needed in order to allow determining the exponential parameters. 
Additionally, the longer time interval is necessary to precisely measure pre-glitch $\ddot{\nu}$ and attempt to detect its change after the glitch.
Such changes are often observed in the presence of recoveries \citep[e.g.][]{cdk88}. 
The investigated interval lies between MJD 51358 (about $67$\,d before the glitch) and MJD 51556 (131\,d after the glitch). 
We avoided using more data before the glitch to prevent the need for a third frequency derivative in the pre-glitch timing model, and we cannot add more data after the event because the recovery is interrupted by the 2000 glitch.

By studying the evolution of $\nu$ and $\dot{\nu}$ in the same time window, we preliminary infer that the main timescale of the recovery is close to $30$\,d.
The DM varies across this time, and we used a linear function to model its behaviour, which gives residuals smaller than $0.005$\,pc\,cm$^{-3}$ in general (Fig.~\ref{fg:glit2}).
The rotational model (with $\tau_\textrm{d}=30$\,d) in combination with the linear DM model were fitted to the high-cadence TOAs at 635 and 950\,MHz. 
This DM model was then kept fixed during the next fits.

The daily TOAs were then used to explore different timescales for the recovery.
Values between $10$ and $50$\,d (in steps of $5$\,d) were tested, and we find that  $30$\,d offers the smallest residuals.
A more detailed exploration around this value shows that $\tau_\textrm{d}=31\pm1$\,d.
The error bar comes from the fact that smaller variations produce negligible differences in the residuals RMS$_w$.
Finally, we used the same process as described above (for the other new glitch) to determine the glitch epoch, with the difference that $\Delta\ddot{\nu}_\textrm{p}$ and $\Delta\nu_\textrm{d}$ were also varied. 
The final best-fit parameters for this glitch are listed in Table \ref{tb:glit2}, and the phase residuals are shown in Fig.~\ref{fg:glit2res}.

\begin{table}
\caption{Parameters measured for the new 1999 glitch and its recovery.}
\label{tb:glit2}
\centering
\begin{tabular}{ll}
\hline\hline
Parameter                                             & Value      \\
\hline
Time range (MJD)    &  51358.18-51555.65\\
Epoch $t_0$ (MJD)    &  51361.00  \\
$\nu$ (Hz)   &  11.1948811980(1) \\
$\dot{\nu}$ ($10^{-15}$\,Hz\,s$^{-1}$)    &  -15573.8(1) \\
$\ddot{\nu}$ ($10^{-21}$\,Hz\,s$^{-2}$)    &  1.88(3) \\
DM$_0$ (pc\,cm$^{-3}$)   &  68.0181(1) \\
DM$_1$ (pc\,cm$^{-3}$\,yr$^{-1}$)   &  -0.0500(5) \\
Glitch epoch (MJD)     & 51425.12(1) \\
$\Delta\nu_\textrm{p}$ ($\mu$Hz)   &  0.251(3) \\
$\Delta\dot{\nu}_\textrm{p}$ ($10^{-15}$\,Hz\,s$^{-1}$)   &  8(1) \\
$\Delta\ddot{\nu}_\textrm{p}$ ($10^{-21}$\,Hz\,s$^{-2}$)   &  -2.6(1) \\
$\Delta\nu_d$ ($\mu$Hz)   &  0.174(3) \\
$\tau_d$ (d)   &  31(1) \\
RMS$_w$ ($\mu$s)     &  25.3 \\
Number of TOAs     &  132 \\
\hline
  {\sl Derived parameters:} & \\
  $\Delta\nu$ ($\mu$Hz)   &  0.425(4) \\
  $\Delta\dot{\nu}$ ($10^{-15}$\,Hz\,s$^{-1}$)   &  -57(3) \\
  $\Delta\ddot{\nu}$ ($10^{-21}$\,Hz\,s$^{-2}$)   &  21.7(1) \\
\hline
\end{tabular}
\tablefoot{1-$\sigma$ error bars are in parentheses and represent the uncertainty of the last quoted digit.
The DM is modelled with the linear function $\rm{DM}(t)=\rm{DM}_0+\rm{DM}_1(t-t_0)$.
}
\end{table}

\section{Timing noise}
\label{sct:tn}
\begin{figure*}
        \centering
        \includegraphics[width=6.0cm]{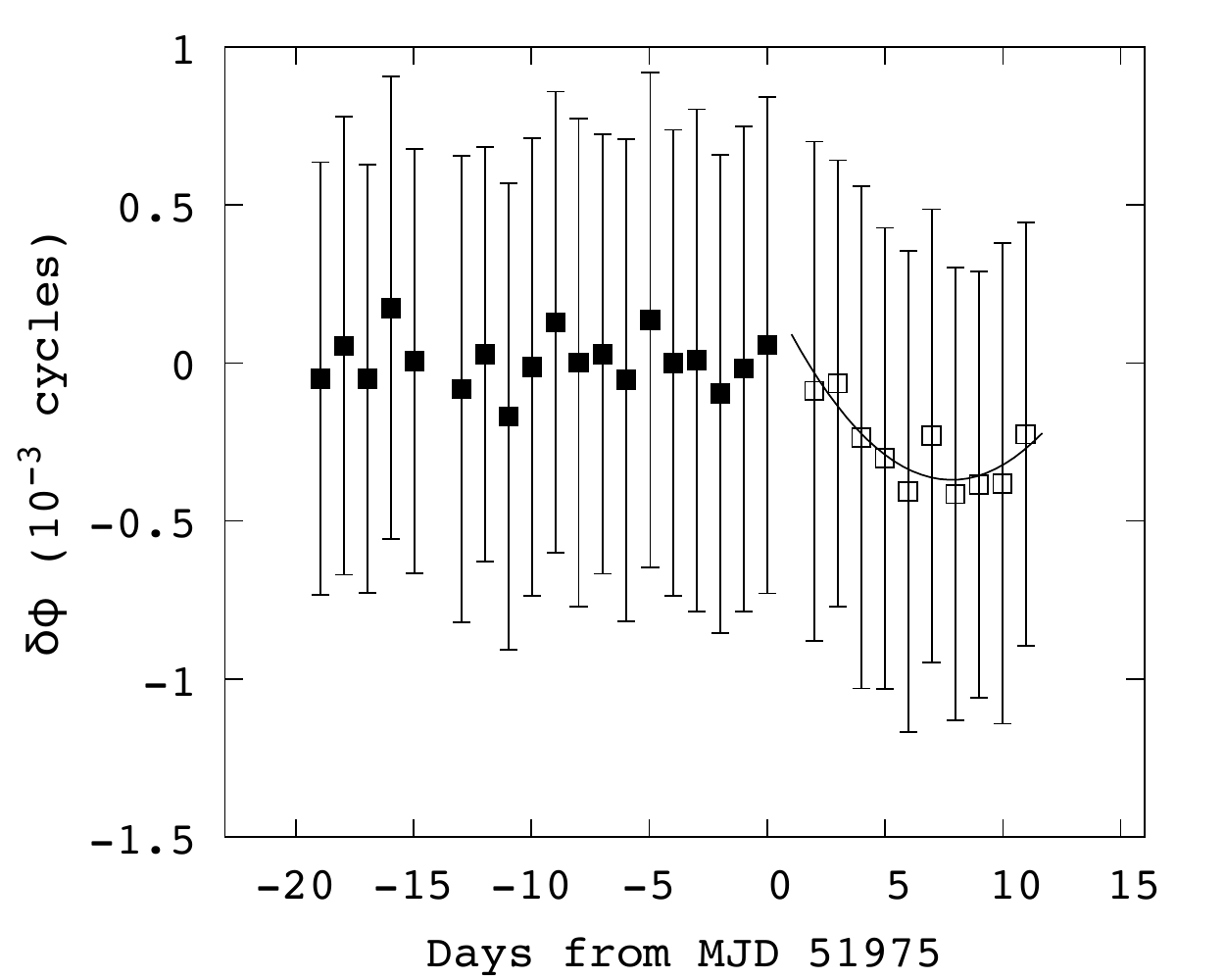}
        \includegraphics[width=6.0cm]{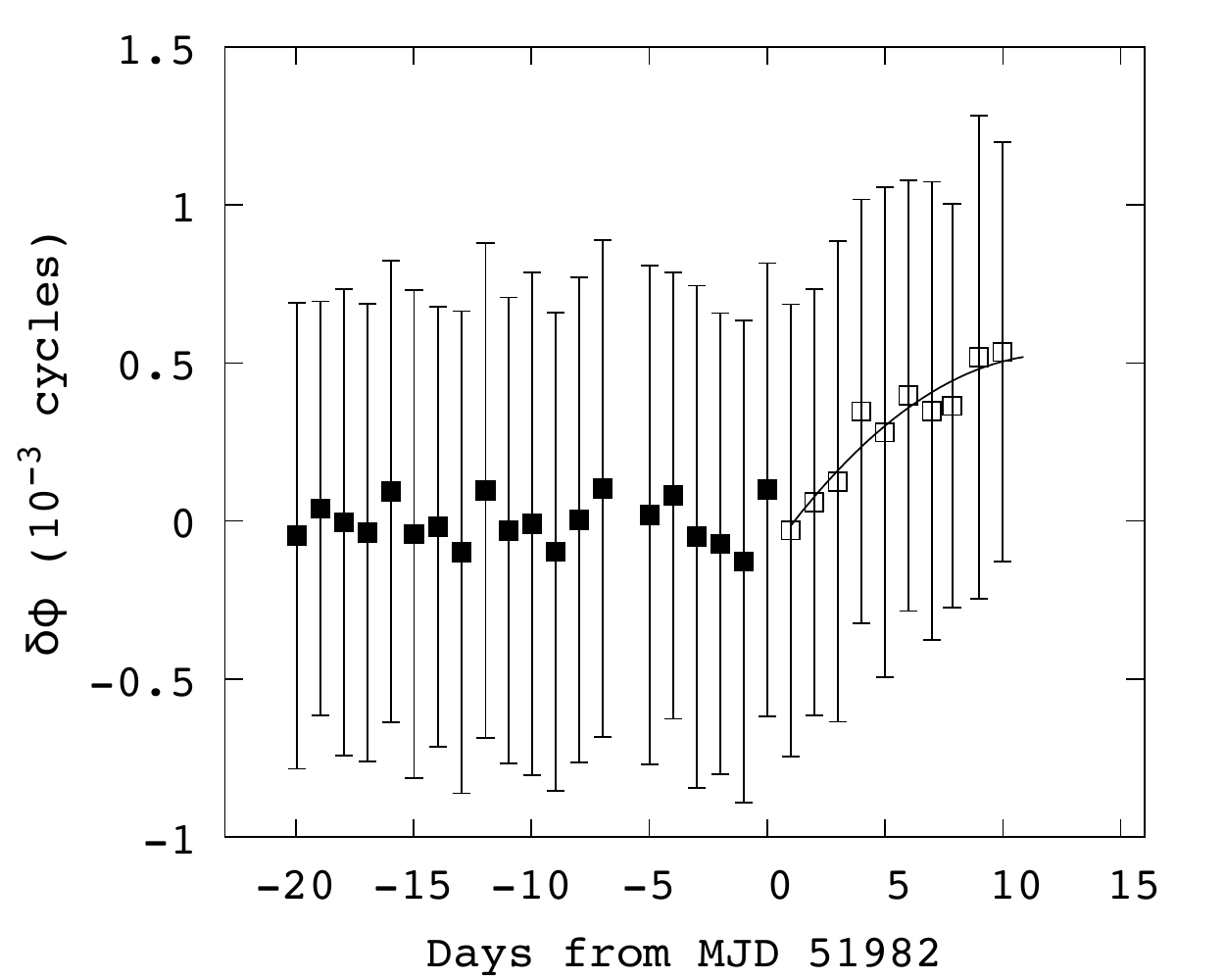}
        \includegraphics[width=6.0cm]{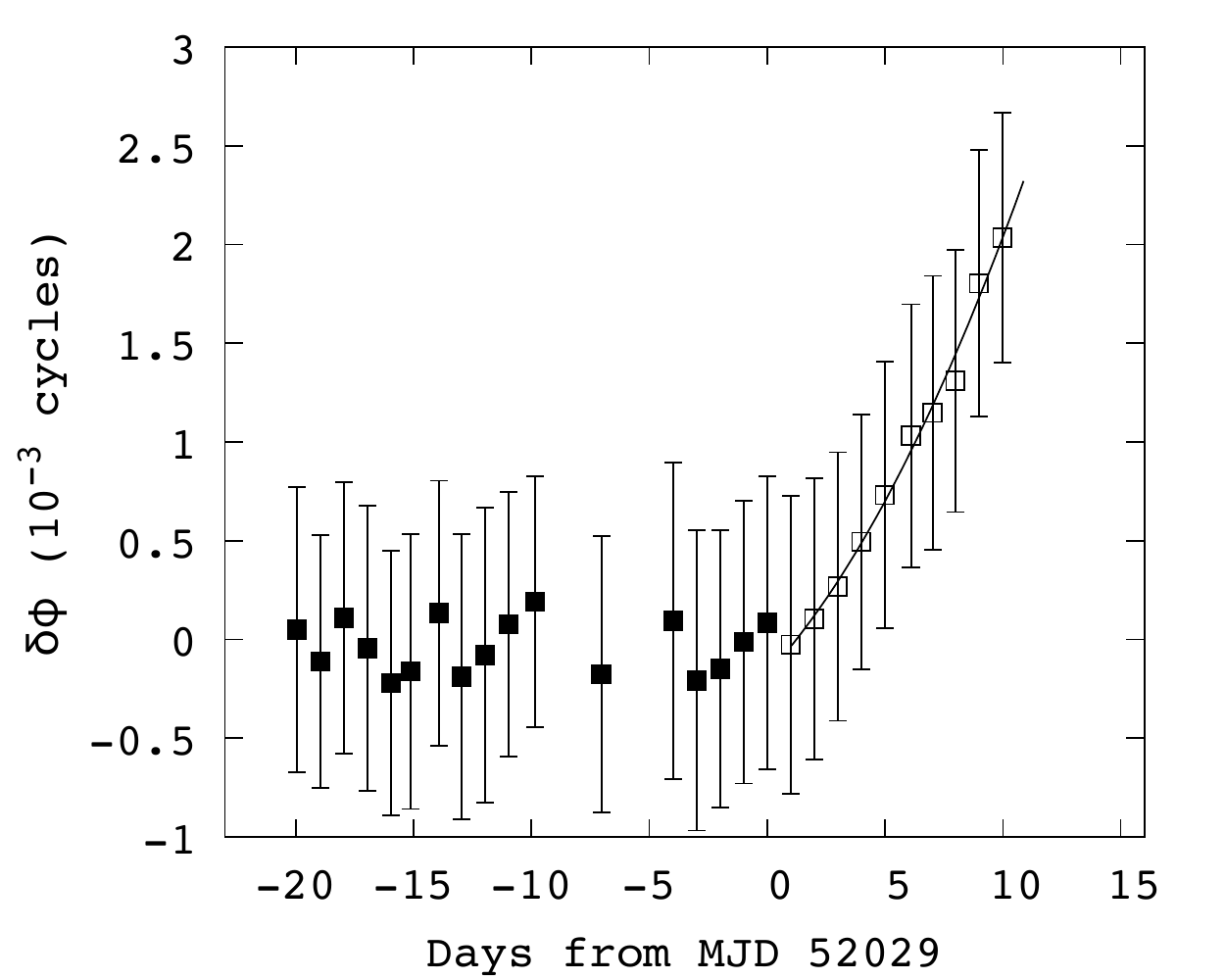}   \\
        \includegraphics[width=6.0cm]{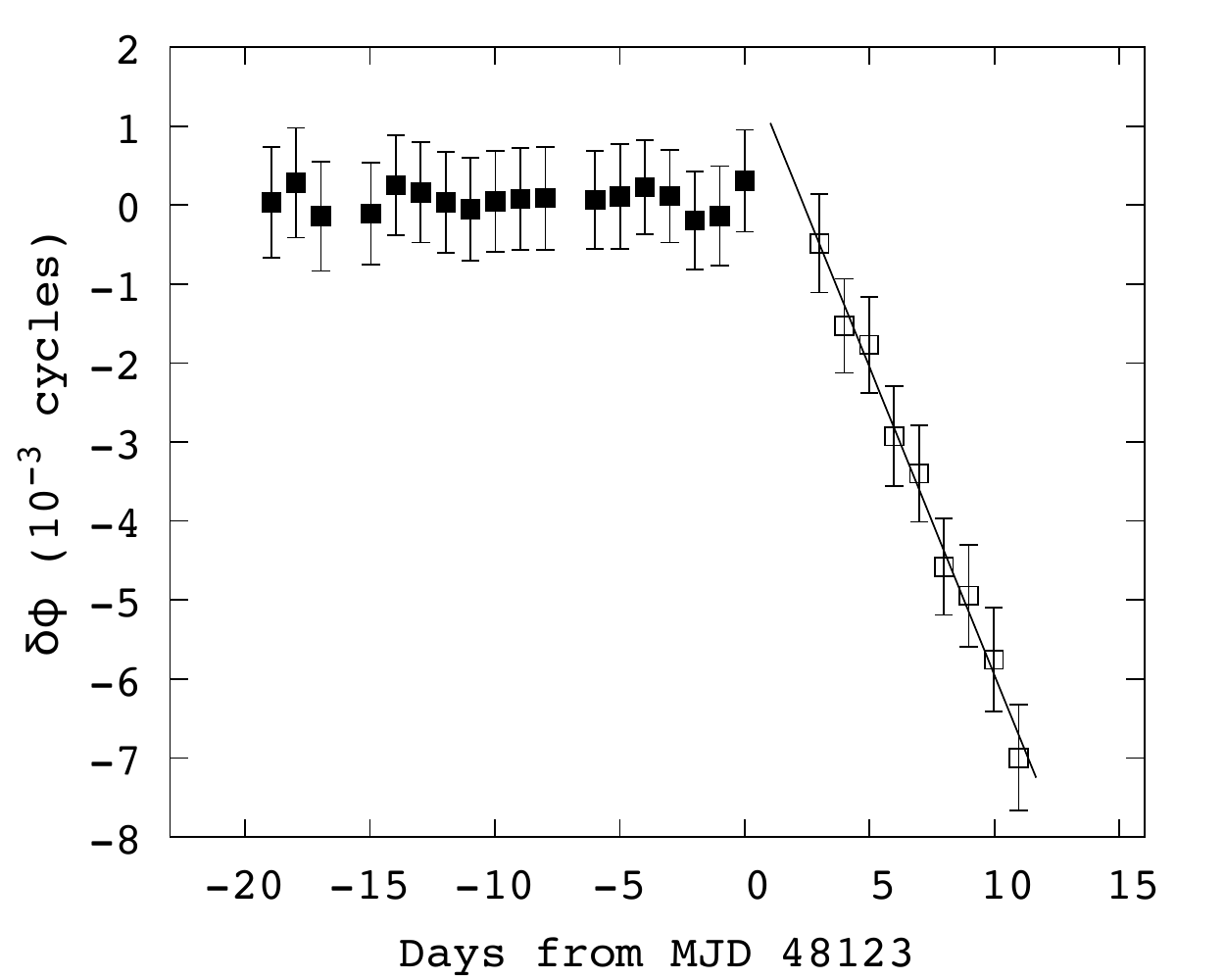} 
        \includegraphics[width=6.0cm]{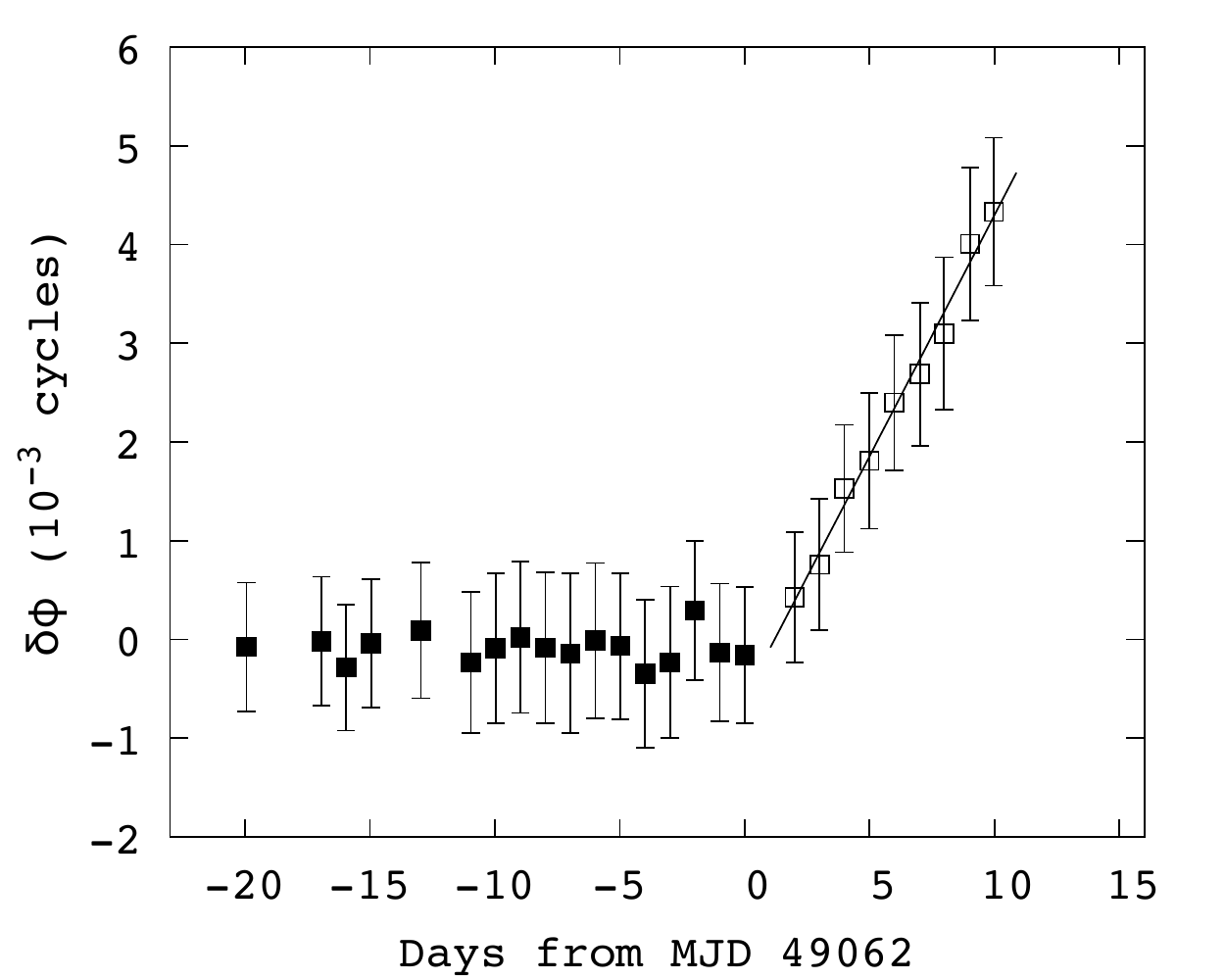}
        \includegraphics[width=6.0cm]{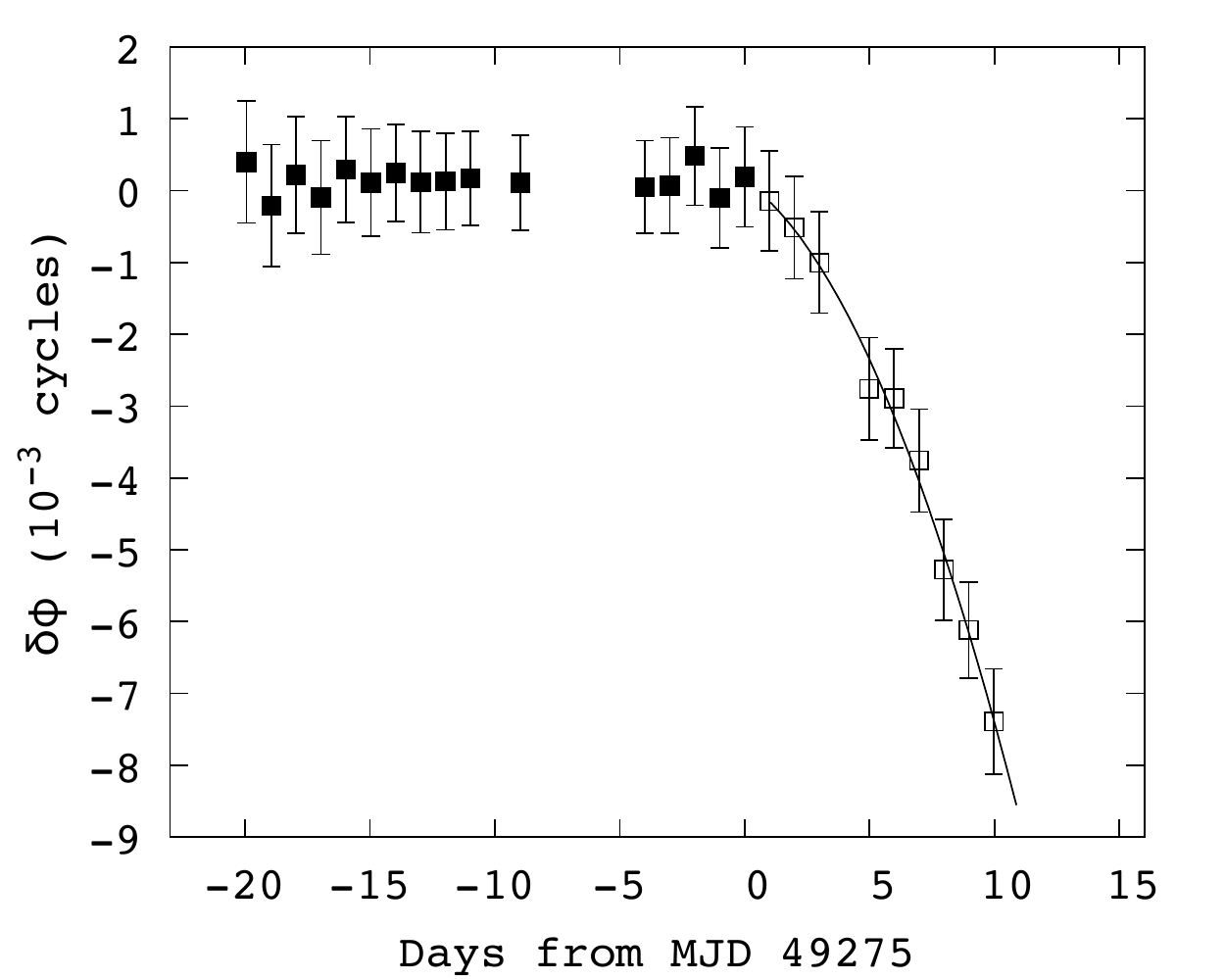} 
        \caption{Examples of two GCs ($\Delta\nu>0$, $\Delta\dot{\nu}<0$; leftmost panels), two AGCs ($\Delta\nu<0$, $\Delta\dot{\nu}>0$; centre), and two events in which $\Delta\nu=0$ and $\Delta\dot{\nu}>0$ (top right) or $\Delta\dot{\nu}<0$ (bottom right).
        The phase residuals are plotted with respect to a fit of TOAs up to the event epoch $t=0$.
        The phase residuals for the \emph{\textup{first}} set are shown with filled squares, and for the \emph{\textup{second}} set, they are shown with empty squares.
        The continuous lines represent the best model for the second set (Eq. \ref{glitmod}).
        For the two cases in the left (left and middle columns) the event at the bottom has a $|\Delta\nu|$ that is much larger than the case at the top.
        Similarly, for the third column, the case at the bottom has a much larger $|\Delta\dot{\nu}|$ than the case at the top.
        }
        \label{res3}
\end{figure*}

The detections presented above correspond to irregularities in the rotation rate that could be modelled as changes $\Delta\nu$  accompanied by changes $\Delta\dot{\nu}$ of the opposite sign, or zero.
However, during the searches, we also identified events with sign combinations of $\Delta\nu$ and $\Delta\dot{\nu}$ that did not correspond to either glitches or anti-glitches, that is, $\Delta\nu>0$ together with $\Delta\dot{\nu}>0$, or $\Delta\nu<0$ together with $\Delta\dot{\nu}<0$.
We also flagged events for which the best-fit involved only a change $\Delta\dot{\nu}$ (i.e.  $\Delta\nu=0$).
It is difficult to confuse the signature of any of these events with glitches.
For example, spin-down-dominated changes produce gradual changes of the slope in the phase residual, rather than cuspy features with rapid changes in slope.
Furthermore, events with changes ($\Delta\nu$, $\Delta\dot{\nu}$) of equal sign produce residual curves  whose slopes increase permanently (becoming either more negative or more positive), as opposed to glitches and anti-glitches (see Fig. \ref{res3}). 
We found 34 events with positive ($\Delta\nu$, $\Delta\dot{\nu}$) changes, 47 events with negative changes, and 67 events in which only $\Delta\dot{\nu}$ was detected.
Examples of their time series together with some GCs and AGCs are shown in Fig.~\ref{fg:rot1}.

\begin{figure*}
  \centering
  \includegraphics[width=15cm]{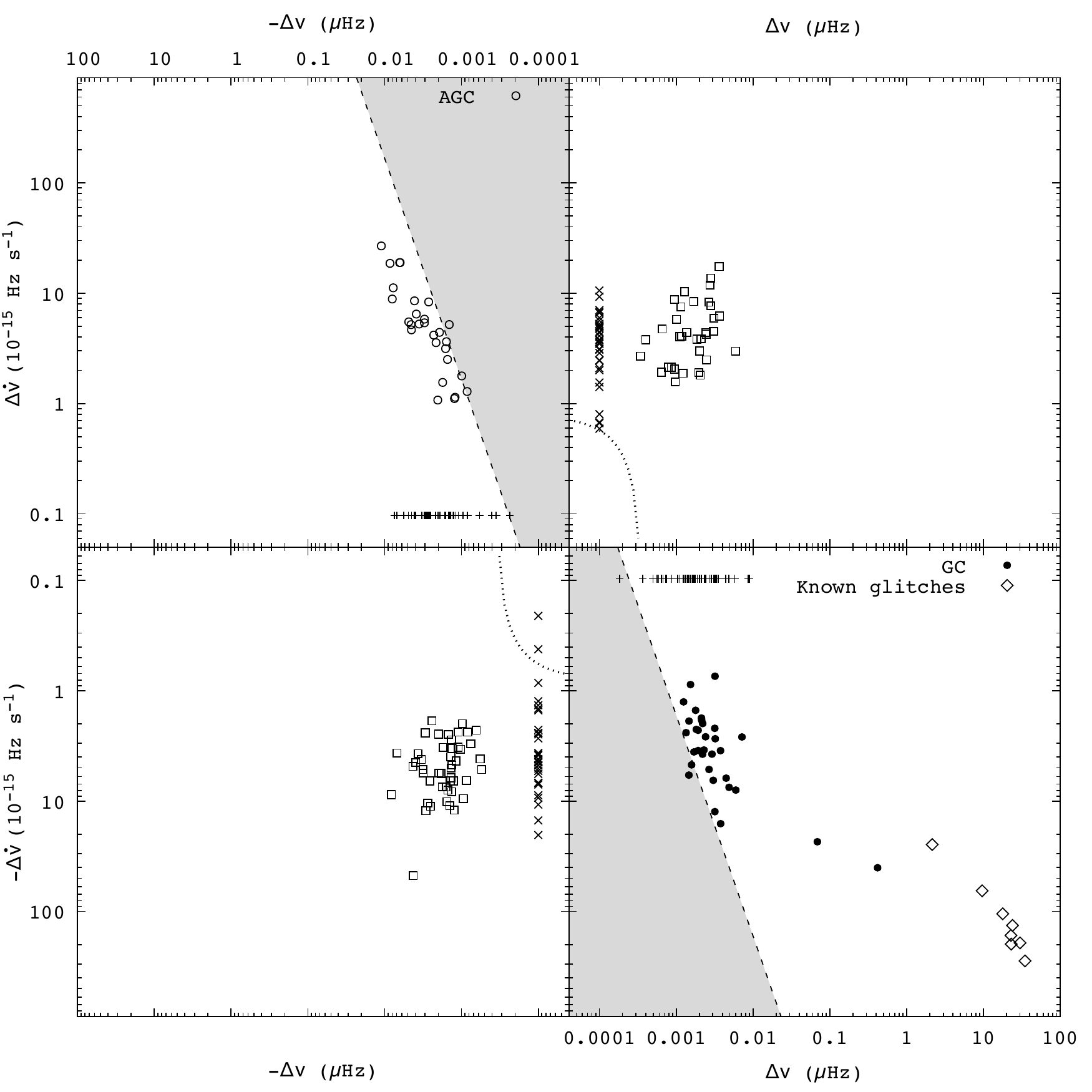}
  \caption{Sizes of all the irregularities selected for all possible combinations of signs for $\Delta\nu$ and $\Delta\dot{\nu}$, including cases in which one of the parameters was not detected ($\Delta\nu=0$ or $\Delta\dot{\nu}=0$, which are plotted as a letter {\sf x} and crosses, respectively).
    These results are for $N_\textrm{d}=10$\,d, and the known Vela glitches in the dataset were measured as for Fig.~\ref{df0df1}.
    The straight lines represent detection limits (events in the shaded areas are unlikely to be detected) when considering average noise levels. 
    The curved dotted lines give an indication of the event sizes below which deviations to the timing model will remain under typical noise levels for $10$\,d since the event epoch (see Section \ref{sct:tn}). 
  }
  \label{fg:rr4}
\end{figure*}

\begin{figure*}
  \centering
  \includegraphics[width=9.1cm]{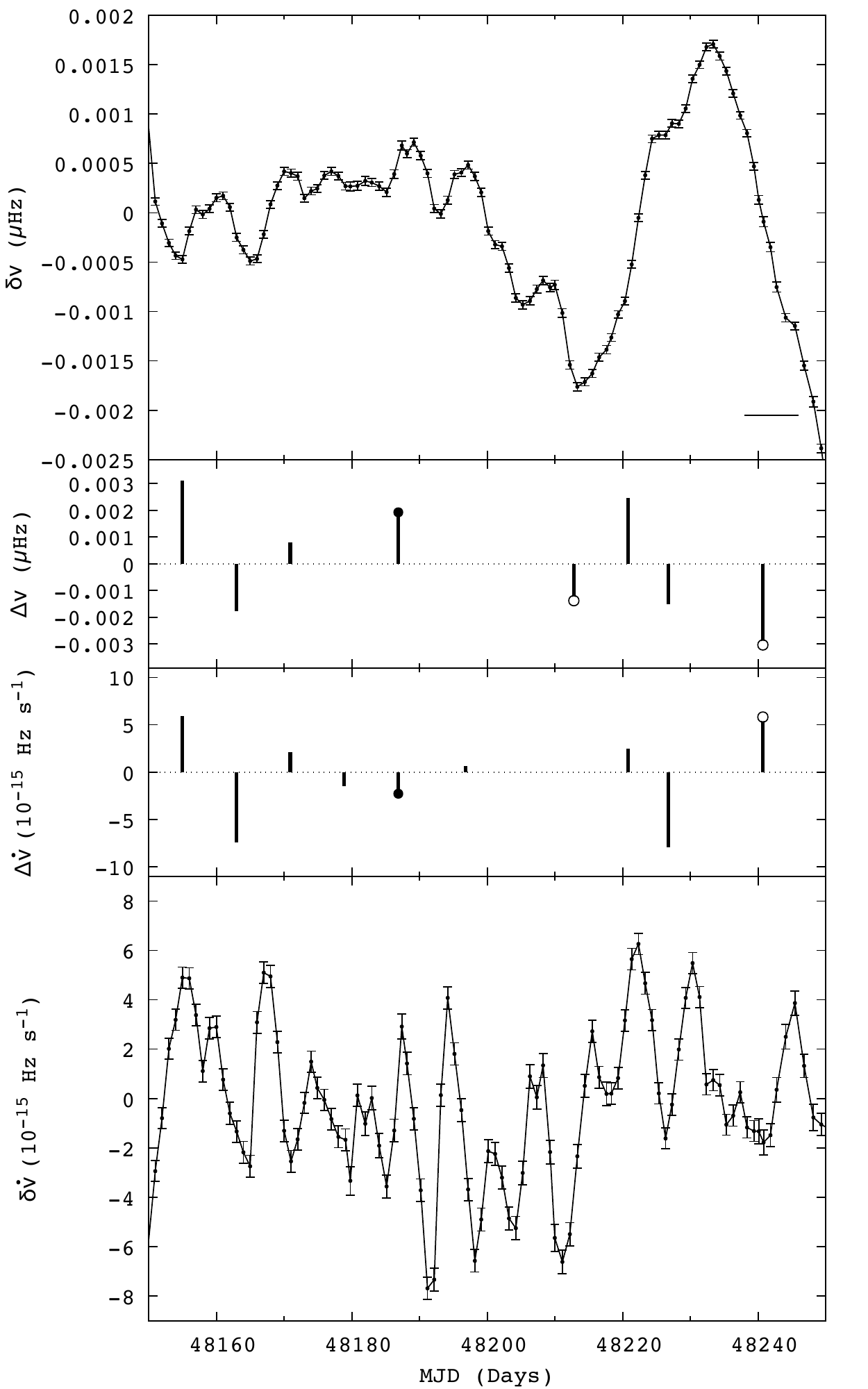}
  \includegraphics[width=9.1cm]{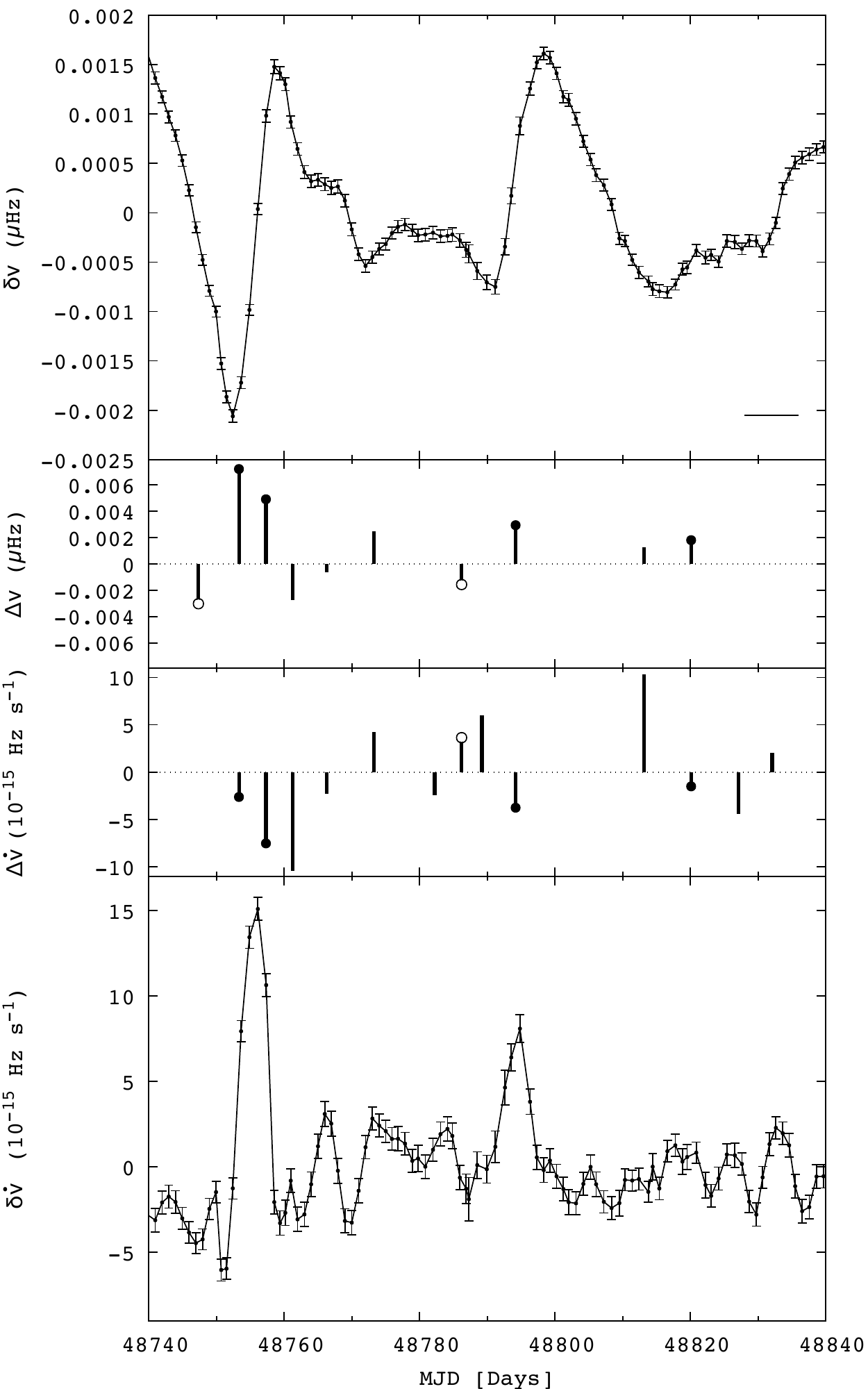}
  \caption{Examples of micro-irregularities in the rotational behaviour of the Vela pulsar and the events selected by the searches.
    {\sl Top:} Frequency residuals relative to a cubic model. 
    {\sl Middle:} Two plots in the centre show the sizes $\Delta\nu$ (top) and $\Delta\dot{\nu}$ (bottom) of the GCs and AGCs (with filled and open circles, respectively, at the end of the bars) and of the other detected irregularities. 
    {\sl Bottom:} Frequency derivative residuals relative to a linear model.
    The $\nu$ and $\dot{\nu}$ datasets were obtained from fits to TOAs in a moving window $8$\,d long that moves by $1$\,d at each stride.
    The length of the horizontal line in the top panel is $8$\,d long.
  }
  \label{fg:rot1}
\end{figure*}

All irregularities selected via the automated searches are shown in Fig.~\ref{fg:rr4}.
We present all possible combinations of signs for $\Delta\nu$ and $\Delta\dot{\nu}$, including GCs and AGCs.
The cases that are best fit with a step change in only one parameter, either $\nu$ or $\dot{\nu}$, are also included in the graph (with small artificial fixed values for the null parameter). 
Most events occupy a specific region in the $\Delta\dot{\nu}$-$\Delta\nu$ space.
For GCs and AGCs we plot the detection limit due to dispersion of the timing residuals $\sigma_\phi^\textrm{max}$ (as explained in the previous sections).
For the other events (quadrants I and III) we used the condition $\phi_\textrm{g}(t=10\,d)>\sigma_\phi^\textrm{max}$ as an indication for event sizes that would be under typical noise levels.

Our results agree with the sizes of \emph{\textup{microglitches}} discussed in \citet{cdk88}.
The results of these authors, like ours, demonstrate that the rotation of the Vela pulsar is interrupted by large glitches approximately every $\sim3$\,yr, whilst the rotation between glitches is affected only by low-level but frequent variations in $\nu$ and $\dot{\nu}$. 
Examples of this behaviour are shown in Fig.~\ref{fg:rot1}, where the evolution of $\nu$ and $\dot{\nu}$ over $100\,$d is shown in detail.
In general, the background rotational activity is dominated by changes in spin-down rate with amplitudes of about a few $10^{-15}$Hz\,s$^{-1}$, although larger fluctuations do sometimes occur.


We used a simple approach to explore whether simulated data with red timing noise contain features that resemble GCs, AGCs, or other events such as those identified by our method. 
Part of the timing noise can be quantified by characterising the spectral properties of the timing residuals using power-law models \citep[e.g.][]{hlk10,slk+16,psj+19}.
We used the \emph{autoSpectralFit} plugin for {\sc tempo2} to measure the spectral properties of Vela residuals of several sections of the daily TOAs.

The power spectra are modelled with a function of the form
\begin{equation}
P(f) = \frac{A}{\left[1+\left(\frac{f}{f_c}\right)^2  \right]^{\alpha/2}} \quad, 
\end{equation}
where $f$ is the spectral frequency, $f_c$ is set to the inverse of the observing time span, $\alpha$ is the spectral index, and $A$ is the amplitude. 
The last two parameters were obtained from a fit to the power spectrum of the real data. 
We then used the \emph{simRedNoise} plugin to simulate data with the exact same cadence, time span, and spectral properties. 
Only inter-glitch sections of data starting at least $\sim100$\,d  post-glitch that span hundreds of days without long TOA gaps were analysed.
The spectral parameters we obtained vary between the analysed data sets. 
Those that best reproduce the observations, in terms of residuals RMS for several time windows and the shape of the residuals features, are those that involve less power at high frequencies (in general, power-law indices $\alpha>3$).
Our choice of parameters to represent the typical inter-glitch residuals of Vela was $\alpha=4.5$, $f_c=0.4$\,yr$^{-1}$, and $A=9.5\times10^{-20}$\,yr$^{3}$.
We present this only as a basic first demonstration, and it is by no means an attempt to model the timing noise in Vela.

Three long series ($\leq10$\,yr) of approximately daily TOAs with these spectral properties were simulated using the \emph{fake} and \emph{simRedNoise} {\sc tempo2} plugins. The simulations use the typical $\nu$, $\dot{\nu}$, $\ddot{\nu}$, and error bar values of Vela. 
The simulated datasets were explored for small rotational step-like changes using the same automated method as we used to analyse the Vela data, and the same selection criteria.
Several timing events can be modelled by steps $\Delta\nu$ and/or $\Delta\dot{\nu}$ of all signs, which appear with a similar frequency as those detected in the real data.
The step sizes of the candidates are in the same range as those found for Vela data ($|\Delta\nu|\leq0.01\,\mu$Hz), and the only difference is that the events in the simulations sometimes involve $\Delta\dot{\nu}$ steps much larger than those observed in the real data.

From this simple exercise, we conclude that the rotation of a Vela-like pulsar affected by noise as simulated here, sampled by daily TOAs, could contain occasional step-like changes $\Delta\nu$ and/or $\Delta\dot{\nu}$. 
However, a more careful and in-depth study is necessary to model timing noise and draw conclusions with regard to its nature in the Vela pulsar.
While this result suggests that the continuous wandering of the rotational phase of a pulsar can mimic step-like rotational changes in timing observations, it does not rule out the existence of such discrete events in the rotation of Vela.

\section{Discussion}

We used nearly continuous observations of the Vela pulsar between 1981 and 2005, which amount to a total of 16.9 years of timing data (four periods lack observations). 
Our analysis complements and significantly extends that of \citet{cdk88}, which covered the glitch activity of Vela from shortly after its discovery up to the recovery of the large glitch at MJD 45192, which is approximately $1.5$ year after the start of our dataset. 
The two studies combined reveal the deviations from a simple rotational model in the course of 37 years, from November 1968 to October 2005.

We have identified small rotational features with amplitudes  $2\times 10^{-4}\leq |\Delta\nu| \leq 10^{-2}\,\mu$Hz
 and $0.4 \leq | \Delta\dot{\nu} | \leq 27\, \times 10^{-15}\,$Hz\,s$^{-1}$ of unknown nature, which contribute to the timing noise of the pulsar. 
\citet{cdk88} described the same phenomenon, and although they used a different approach to characterise these irregularities, their results agree with those presented here. 
Even though it cannot be excluded that some of these features are of the same origin as the large glitches, there does not appear to be a continuum of typical glitches from the large ($\Delta\nu>10\,\mu$Hz) glitches down to the small-scale events ($\Delta\nu< 10^{-2}\,\mu$Hz; see Fig. \ref{histos}). 
Moreover, the noise-like irregularities are often gradual and more resemble a torque-driven process such as described in \citet{lhk+10}. 
Fig.~\ref{fg:rot1} shows that a change in spin-down rate sometimes clearly precedes the changes in rotation rate. 
Even though we used discrete steps to describe and quantify part of the timing noise, the underlying phenomenon is perhaps not just a mere composition of individual events.
Instead, we might be detecting the sharpest changes of a process that in general involves gradual deviations of the parameters. 
We reserve a more refined analysis of timing noise in the Vela pulsar, including variability on longer timescales, for a separate study.


When the entire pulsar population is considered, the overall distribution of glitch sizes appears bimodal.
While the majority distribute broadly from very small sizes up to $1$-$10\,\mu$Hz, the large glitches are narrowly clustered around $\Delta\nu\simeq20\,\mu$Hz \citep{ymh+13,ka14b,apj17,fer+17}. 
The glitches of the Vela pulsar belong mostly to this narrow second peak, and we have shown here that the deficit of smaller glitches is not an artefact of data limitations or observational analyses.

A few other pulsars exhibit mostly large glitches, with varying rates of occurrence. Some examples are PSRs B1757$-$24, B1800$-$21, and B1823$-$13 \citep{elsk11,els17}. 
Not all young pulsars behave like this, however.
Others display a majority of small glitches, with the occasional larger event \citep[e.g.][for the Crab pulsar]{sls+18}, showing power-law size distributions and highly irregular waiting times between events \citep[e.g. PSRs J0631+1036 and B1737$-$30;][]{hmd18,fer19}.
This difference in pulsar glitching behaviour constitutes one of the main open questions in glitch physics \citep{mpw08,has16}. 
It is a reasonable speculation that perhaps a different mechanism operates for small and large glitches, or that they are driven from a separate superfluid reservoir altogether \citep{hkaa18}.

Twenty-three glitches are known for Vela (including the two new glitches presented here)\footnote{Glitch sizes and epochs were taken from the Jodrell Bank glitch catalogue: \url{http://www.jb.man.ac.uk/pulsar/glitches.html}, except for the two new glitches found in this work}.
Of these, 17 occurred in the 37 years mentioned above. This list is possibly complete for glitches of $\Delta\nu$ greater than $\sim0.001\,\mu$Hz.
During this time, only three glitches were smaller than $1\,\mu$Hz and 14 glitches were larger \citep[the small ones are the 2 new glitches presented here and the small glitch reported by][]{cdk88}.
In the following $\sim13$\,yr,  5 large glitches and only one small glitch were reported, whose tiny size \citep[$\Delta\nu=0.004\,\mu$Hz,][]{jbb+15} suggests that it might be part of the population of events that contribute to the timing noise instead of a typical glitch \citep[as indicated by][]{lbs+20}.

This shows that the occurrence rate of small glitches ($\Delta\nu<1\,\mu$Hz) is very low in the Vela pulsar.
The two small events detected in $16.9$\,yr lead to an occurrence rate of small glitches of about $0.1$\,yr$^{-1}$ (one every 8.5 years).  
This is consistent with the rate implied by the results of \citet[][one small glitch in $14$\,yr]{cdk88}.
According to this rate, the possibility that a small glitch is missed in any of the three gaps within the data examined in this work is marginal (an expected $0.8$ glitches in the accumulated $7.1$\,yr of the three gaps).
On the other hand, the rate of large glitches ($\Delta\nu>1\,\mu$Hz) is much higher: $0.4$\,yr$^{-1}$ (i.e. one every $2.6$\,yr).

The above limit of $1\,\mu$Hz was chosen somewhat arbitrarily, but it is worth noting that only two  glitches with sizes between $1$ and $10\,\mu$Hz are reported (in all $37+13$\,yr of Vela monitoring). 
These two events occurred very close to each other, only $32\,$d apart, and their added sizes amount to $11.6\,\mu$Hz \citep[e.g.][]{bf11}.

The Vela distribution of glitch sizes ($\Delta\nu$) is usually described by a Gaussian function \citep{mpw08,hmd18,fer19}. 
For the 17-glitches sample, a KS test returns a modest probability that their sizes follow a normal distribution ($p_\textrm{KS}=0.60$). 
We confirmed that a variety of other probability distribution functions (PDFs) led to worse results.
For a subset of only the largest glitches ($\Delta\nu>1\,\mu$Hz), the fit to a Gaussian only slightly improves, with the new $p_\textrm{KS}=0.75$. 
Using the best-fit parameters for a Gaussian PDF, we calculate that the probability of drawing a glitch with a size equal to the second largest GC ($\Delta\nu=0.0621\pm0.0003\,\mu$Hz, Table \ref{tb:glit1}), which we believe is in fact a glitch, is $2\times10^{-6}$ when all 17 glitches are considered, or $7\times10^{-7}$ when the smallest events are excluded. 
This is low, but still one to two orders of magnitude larger than the probability that this event belongs to the GC best-fit log-normal distribution.

In addition to their size and the occurrence rate, another property differs between small and large Vela glitches: the latter appear somewhat quasi-periodically \citep{mpw08,fer19,cm19}, with an average separation of about 1,000\,d (and a standard deviation $\sim300$\,d), whilst small glitches seem to occur randomly. 
Lastly, another possible separating factor between small and large glitches is the linear correlation between the size of a glitch and the time to the next one. 
This correlation has so far been observed with high significance only for the X-ray pulsar PSR~J0537$-$6910 \citep[Spearman's rank correlation coefficient $r_s=0.95$; e.g.][]{mgm+04,aeka18}. 
The question now is whether glitch size and time-to-the-next-glitch are correlated in the Vela pulsar.

To try to answer this question, we examined only the time period for which the glitch sample is probably complete, and therefore we can be more certain for the inferred inter-glitch waiting times.   
With this sample (17 glitches in 37 years) we did not find evidence for a correlation  ($r_s = 0.19$ and probability $p_s = 0.47$). 
However, \citet{fer19} found that the correlation in PSR~J0537$-$6910 is tighter when the smaller events (6 out of a total 45 glitches) are not considered, and that a linear fit passes closer to the origin when only the larger events are kept.  
The authors also found some indication in some other pulsars that the correlation improves when only large events are considered \citep[e.g PSR J0205+6449 with $r_s>0.75$; see also][]{mhf18}. 
For Vela, using all known glitches at the time, the authors showed that when the smallest ones $(<2\,\mu$Hz) are excluded, a weak correlation emerges. 
Using the same cut-off as \citet{fer19} and all Vela glitches in the literature, we verified a weak correlation, with $r_s=0.6$ ($p_s = 0.009$). 
In the time range MJD 40280-53193 (where probably no glitches have been missed), and with a less strict cut-off, at $0.4\,\mu$Hz, the correlation further improves with $r_s = 0.7$ and $p_s= 0.004$. This cut-off allows the new glitch, which showed clear relaxation, but leaves out the smallest of the new glitches. 
This provides further support to the idea that larger (typical?) glitches of the Vela pulsar form a different population from smaller events and obey a size-waiting time correlation as in PSR~J0537$-$6910, albeit weaker. Whilst we acknowledge that parts of this discussion might be speculative and of low statistical weight, they offer a direction for future explorations of the nature of different glitching behaviour in neutron stars.

\section{Conclusions}
We demonstrated that there is a real deficit of small glitches ($\Delta\nu<1\,\mu$Hz) in the Vela pulsar, both in comparison to its numerous large glitches and to the frequency of small glitches seen in other pulsars such as the Crab.
The rate of glitches in the Vela pulsar is four times lower than that of large glitches.
While the sizes of the large glitches can be described by a Gaussian distribution, it is unclear whether the small glitches belong to the same distribution or to a different component.
In addition to the glitches, there is a noise-like component that involves events that can be modelled by sudden changes ($\Delta\nu$, $\Delta\dot{\nu}$) of all signs and that involves small frequency changes ($|\Delta\nu|<0.02\,\mu$Hz).

\section{Acknowledgements}
We have made extensive use of the graphing utility GNUPLOT.
C.M.E was supported by the grants ANID FONDECYT/Regular 1171421; and USA1899 - Vridei 041931SSSA-PAP (Universidad de Santiago de Chile, USACH). D.A. acknowledges support from the Polish National Science Centre grant SONATA BIS 2015/18/E/ST9/00577.

\bibliographystyle{aa}
\bibliography{journals,pulsar}

\begin{thebibliography}{57}
\expandafter\ifx\csname natexlab\endcsname\relax\def\natexlab#1{#1}\fi

\bibitem[{{Akbal} \& {Alpar}(2018)}]{aa18}
{Akbal}, O. \& {Alpar}, M.~A. 2018, \mnras, 473, 621

\bibitem[{{Anderson} \& {Itoh}(1975)}]{ai75}
{Anderson}, P.~W. \& {Itoh}, N. 1975, Nature, 256, 25

\bibitem[{Andersson {et~al.}(2012)Andersson, Glampedakis, Ho, \&
  Espinoza}]{aghe12}
Andersson, N., Glampedakis, K., Ho, W. C.~G., \& Espinoza, C.~M. 2012, Phys.
  Rev. Lett., 109

\bibitem[{{Antonopoulou} {et~al.}(2018){Antonopoulou}, {Espinoza}, {Kuiper}, \&
  {Andersson}}]{aeka18}
{Antonopoulou}, D., {Espinoza}, C.~M., {Kuiper}, L., \& {Andersson}, N. 2018,
  MNRAS, 473, 1644

\bibitem[{Aschenbach {et~al.}(1995)Aschenbach, Egger, \& Tr{\"u}mper}]{aet95}
Aschenbach, B., Egger, R., \& Tr{\"u}mper, J. 1995, Nature, 373, 587

\bibitem[{{Ashton} {et~al.}(2017){Ashton}, {Prix}, \& {Jones}}]{apj17}
{Ashton}, G., {Prix}, R., \& {Jones}, D.~I. 2017, Phys. Rev. D, 96, 063004

\bibitem[{{Buchner}(2013)}]{buc13b}
{Buchner}, S. 2013, in IAU Symposium, Vol. 291, Neutron Stars and Pulsars:
  Challenges and Opportunities after 80 years, ed. J.~{van Leeuwen}, 207--207

\bibitem[{{Buchner} \& {Flanagan}(2011)}]{bf11}
{Buchner}, S. \& {Flanagan}, C. 2011, in American Institute of Physics
  Conference Series, Vol. 1357, Radio Pulsars: An Astrophysical Key to Unlock
  the Secrets of the Universe, ed. M.~{Burgay}, N.~{D'Amico}, P.~{Esposito},
  A.~{Pellizzoni}, \& A.~{Possenti}, 113--116

\bibitem[{Carlin \& Melatos(2019)}]{cm19}
Carlin, J.~B. \& Melatos, A. 2019, MNRAS, 483, 4742

\bibitem[{Chamel(2013)}]{cha13}
Chamel, N. 2013, Phys. Rev. Lett., 110, 011101

\bibitem[{Cordes {et~al.}(1988)Cordes, Downs, \& Krause-{P}olstorff}]{cdk88}
Cordes, J.~M., Downs, G.~S., \& Krause-{P}olstorff, J. 1988, ApJ, 330, 847

\bibitem[{{Delsate} {et~al.}(2016){Delsate}, {Chamel}, {G{\"u}rlebeck},
  {Fantina}, {Pearson}, \& {Ducoin}}]{dcg+16}
{Delsate}, T., {Chamel}, N., {G{\"u}rlebeck}, N., {et~al.} 2016, Phys. Rev. D,
  94, 023008

\bibitem[{{Dodson} {et~al.}(2003){Dodson}, {Legge}, {Reynolds}, \&
  {McCulloch}}]{dlrm03}
{Dodson}, R., {Legge}, D., {Reynolds}, J.~E., \& {McCulloch}, P.~M. 2003, ApJ,
  596, 1137

\bibitem[{{Dodson} {et~al.}(2007){Dodson}, {Lewis}, \& {McCulloch}}]{dlm07}
{Dodson}, R., {Lewis}, D., \& {McCulloch}, P. 2007, Astrophys. Space Sci., 308,
  585

\bibitem[{{Dodson} {et~al.}(2002){Dodson}, {McCulloch}, \& {Lewis}}]{dml02}
{Dodson}, R.~G., {McCulloch}, P.~M., \& {Lewis}, D.~R. 2002, ApJ, 564, L85

\bibitem[{{Edwards} {et~al.}(2006){Edwards}, {Hobbs}, \& {Manchester}}]{ehm06}
{Edwards}, R.~T., {Hobbs}, G.~B., \& {Manchester}, R.~N. 2006, MNRAS, 372, 1549

\bibitem[{{Espinoza} {et~al.}(2014){Espinoza}, {Antonopoulou}, {Stappers},
  {Watts}, \& {Lyne}}]{eas+14}
{Espinoza}, C.~M., {Antonopoulou}, D., {Stappers}, B.~W., {Watts}, A., \&
  {Lyne}, A.~G. 2014, MNRAS, 440, 2755

\bibitem[{{Espinoza} {et~al.}(2017){Espinoza}, {Lyne}, \& {Stappers}}]{els17}
{Espinoza}, C.~M., {Lyne}, A.~G., \& {Stappers}, B.~W. 2017, MNRAS, 466, 147

\bibitem[{{Espinoza} {et~al.}(2011){Espinoza}, {Lyne}, {Stappers}, \&
  {Kramer}}]{elsk11}
{Espinoza}, C.~M., {Lyne}, A.~G., {Stappers}, B.~W., \& {Kramer}, M. 2011,
  MNRAS, 414, 1679

\bibitem[{Flanagan(1990)}]{fla90}
Flanagan, C.~S. 1990, Nature, 345, 416

\bibitem[{Fuentes {et~al.}(2019)Fuentes, Espinoza, \& Reisenegger}]{fer19}
Fuentes, J.~R., Espinoza, C.~M., \& Reisenegger, A. 2019, A\&A, 630, A115

\bibitem[{{Fuentes} {et~al.}(2017){Fuentes}, {Espinoza}, {Reisenegger}, {Shaw},
  {Stappers}, \& {Lyne}}]{fer+17}
{Fuentes}, J.~R., {Espinoza}, C.~M., {Reisenegger}, A., {et~al.} 2017, A\&A,
  608, A131

\bibitem[{Hamilton {et~al.}(1985)Hamilton, Hall, \& Costa}]{hhc85}
Hamilton, P.~A., Hall, P.~J., \& Costa, M.~E. 1985, MNRAS, 214, 5{P}

\bibitem[{{Haskell}(2016)}]{has16}
{Haskell}, B. 2016, MNRAS, 461, L77

\bibitem[{{Haskell} {et~al.}(2018){Haskell}, {Khomenko}, {Antonelli}, \&
  {Antonopoulou}}]{hkaa18}
{Haskell}, B., {Khomenko}, V., {Antonelli}, M., \& {Antonopoulou}, D. 2018,
  MNRAS, 481, L146

\bibitem[{{Haskell} \& {Melatos}(2015)}]{hm15}
{Haskell}, B. \& {Melatos}, A. 2015, Int. J. Mod. Phys. D., 24, 30008

\bibitem[{{Haskell} {et~al.}(2012){Haskell}, {Pizzochero}, \& {Sidery}}]{hps12}
{Haskell}, B., {Pizzochero}, P.~M., \& {Sidery}, T. 2012, MNRAS, 420, 658

\bibitem[{Ho {et~al.}(2015)Ho, Espinoza, Antonopoulou, \& Andersson}]{heaa15}
Ho, W.~C.~G., Espinoza, C.~M., Antonopoulou, D., \& Andersson, N. 2015, Science
  Advances, 1, e1500578

\bibitem[{{Hobbs} {et~al.}(2010){Hobbs}, {Lyne}, \& {Kramer}}]{hlk10}
{Hobbs}, G., {Lyne}, A.~G., \& {Kramer}, M. 2010, MNRAS, 402, 1027

\bibitem[{{Hobbs} {et~al.}(2006){Hobbs}, {Edwards}, \& {Manchester}}]{hem06}
{Hobbs}, G.~B., {Edwards}, R.~T., \& {Manchester}, R.~N. 2006, MNRAS, 369, 655

\bibitem[{{Howitt} {et~al.}(2018){Howitt}, {Melatos}, \& {Delaigle}}]{hmd18}
{Howitt}, G., {Melatos}, A., \& {Delaigle}, A. 2018, \apj, 867, 60

\bibitem[{{Jankowski} {et~al.}(2015){Jankowski}, {Bailes}, {Barr}, {Bateman},
  {Bhandari}, {Briggs}, {Caleb}, {Campbell-Wilson}, {Flynn}, {Green},
  {Hunstead}, {Jameson}, {Keane}, {Ravi}, {Krishnan}, \& {van
  Straten}}]{jbb+15}
{Jankowski}, F., {Bailes}, M., {Barr}, E., {et~al.} 2015, The Astronomer's
  Telegram, 6903, 1

\bibitem[{{Konar} \& {Arjunwadkar}(2014)}]{ka14b}
{Konar}, S. \& {Arjunwadkar}, M. 2014, in Astronomical Society of India
  Conference Series, Vol.~13, 87--88

\bibitem[{{Kramer} {et~al.}(2006){Kramer}, {Lyne}, {O'Brien}, {Jordan}, \&
  {Lorimer}}]{klo+06}
{Kramer}, M., {Lyne}, A.~G., {O'Brien}, J.~T., {Jordan}, C.~A., \& {Lorimer},
  D.~R. 2006, Science, 312, 549

\bibitem[{Large {et~al.}(1968)Large, Vaughan, \& Mills}]{lvm68}
Large, M.~I., Vaughan, A.~E., \& Mills, B.~Y. 1968, Nature, 220, 340

\bibitem[{{Link} {et~al.}(1999){Link}, {Epstein}, \& {Lattimer}}]{lel99}
{Link}, B., {Epstein}, R.~I., \& {Lattimer}, J.~M. 1999, Phys. Rev. Lett., 83,
  3362

\bibitem[{Lorimer \& Kramer(2005)}]{lk05}
Lorimer, D.~R. \& Kramer, M. 2005, Handbook of Pulsar Astronomy (Cambridge
  University Press)

\bibitem[{Lower {et~al.}(2020)Lower, Bailes, Shannon, Johnston, Flynn,
  Os{\l}owski, Gupta, Farah, Bateman, Green, Hunstead, Jameson, Jankowski,
  Parthasarathy, Price, Sutherland, Temby, \& Venkatraman~Krishnan}]{lbs+20}
Lower, M.~E., Bailes, M., Shannon, R.~M., {et~al.} 2020, MNRAS, 494, 228

\bibitem[{{Lyne} {et~al.}(2010){Lyne}, {Hobbs}, {Kramer}, {Stairs}, \&
  {Stappers}}]{lhk+10}
{Lyne}, A., {Hobbs}, G., {Kramer}, M., {Stairs}, I., \& {Stappers}, B. 2010,
  Science, 329, 408

\bibitem[{{Marshall} {et~al.}(2004){Marshall}, {Gotthelf}, {Middleditch},
  {Wang}, \& {Zhang}}]{mgm+04}
{Marshall}, F.~E., {Gotthelf}, E.~V., {Middleditch}, J., {Wang}, Q.~D., \&
  {Zhang}, W. 2004, ApJ, 603, 682

\bibitem[{Melatos {et~al.}(2018)Melatos, Howitt, \& Fulgenzi}]{mhf18}
Melatos, A., Howitt, G., \& Fulgenzi, W. 2018, ApJ, 863, 196

\bibitem[{Melatos \& Link(2014)}]{ml14}
Melatos, A. \& Link, B. 2014, MNRAS, 437, 21

\bibitem[{{Melatos} {et~al.}(2008){Melatos}, {Peralta}, \& {Wyithe}}]{mpw08}
{Melatos}, A., {Peralta}, C., \& {Wyithe}, J.~S.~B. 2008, ApJ, 672, 1103

\bibitem[{Melatos \& Warszawski(2009)}]{mw09}
Melatos, A. \& Warszawski, L. 2009, ApJ, 700, 1524

\bibitem[{Palfreyman {et~al.}(2018)Palfreyman, Dickey, Hotan, Ellingsen, \&
  Straten}]{pdh+18}
Palfreyman, J., Dickey, J.~M., Hotan, A., Ellingsen, S., \& Straten, W. 2018,
  Nature, 556, 219

\bibitem[{Palfreyman {et~al.}(2016)Palfreyman, Dickey, Ellingsen, Jones, \&
  Hotan}]{pde+16}
Palfreyman, J.~L., Dickey, J.~M., Ellingsen, S.~P., Jones, I.~R., \& Hotan,
  A.~W. 2016, ApJ, 820, 64

\bibitem[{Parthasarathy {et~al.}(2019)Parthasarathy, Shannon, Johnston,
  Lentati, Bailes, Dai, Kerr, Manchester, Os{\l}owski, Sobey, van Straten, \&
  Weltevrede}]{psj+19}
Parthasarathy, A., Shannon, R.~M., Johnston, S., {et~al.} 2019, MNRAS, 489,
  3810

\bibitem[{Petroff {et~al.}(2013)Petroff, Keith, Johnston, van Straten, \&
  Shannon}]{pkj+13}
Petroff, E., Keith, M.~J., Johnston, S., van Straten, W., \& Shannon, R.~M.
  2013, MNRAS, 435, 1610

\bibitem[{Pizzochero {et~al.}(2017)Pizzochero, Antonelli, Haskell, \&
  Seveso}]{pahs17}
Pizzochero, P.~M., Antonelli, M., Haskell, B., \& Seveso, S. 2017, Nature
  Astronomy, 1, 1

\bibitem[{Radhakrishnan \& Manchester(1969)}]{rm69}
Radhakrishnan, V. \& Manchester, R.~N. 1969, Nature, 222, 228

\bibitem[{Reichley \& Downs(1969)}]{rd69}
Reichley, P.~E. \& Downs, G.~S. 1969, Nature, 222, 229

\bibitem[{Shannon {et~al.}(2016)Shannon, Lentati, Kerr, Johnston, Hobbs, \&
  Manchester}]{slk+16}
Shannon, R.~M., Lentati, L.~T., Kerr, M., {et~al.} 2016, MNRAS, 459, 3104

\bibitem[{{Shaw} {et~al.}(2018){Shaw}, {Lyne}, {Stappers}, {Weltevrede},
  {Bassa}, {Lien}, {Mickaliger}, {Breton}, {Jordan}, {Keith}, \&
  {Krimm}}]{sls+18}
{Shaw}, B., {Lyne}, A.~G., {Stappers}, B.~W., {et~al.} 2018, {MNRAS}, 478, 3832

\bibitem[{Shemar \& Lyne(1996)}]{sl96}
Shemar, S.~L. \& Lyne, A.~G. 1996, MNRAS, 282, 677

\bibitem[{{Sushch} \& {Hnatyk}(2014)}]{sh14}
{Sushch}, I. \& {Hnatyk}, B. 2014, A\&A, 561, A139

\bibitem[{{Tsuruta} {et~al.}(2009){Tsuruta}, {Sadino}, {Kobelski}, {Teter},
  {Liebmann}, {Takatsuka}, {Nomoto}, \& {Umeda}}]{tsk+09}
{Tsuruta}, S., {Sadino}, J., {Kobelski}, A., {et~al.} 2009, ApJ, 691, 621

\bibitem[{{Yu} {et~al.}(2013){Yu}, {Manchester}, {Hobbs}, {Johnston}, {Kaspi},
  {Keith}, {Lyne}, {Qiao}, {Ravi}, {Sarkissian}, {Shannon}, \& {Xu}}]{ymh+13}
{Yu}, M., {Manchester}, R.~N., {Hobbs}, G., {et~al.} 2013, MNRAS, 429, 688

\end{thebibliography}

\end{document}